\documentclass[lettersize,journal]{IEEEtran}
\usepackage{amsmath,amsfonts}
\usepackage{algorithmic}
\usepackage{array}
\usepackage[caption=false,font=normalsize,labelfont=sf,textfont=sf]{subfig}
\usepackage{textcomp}
\usepackage{stfloats}
\usepackage{url}
\usepackage{verbatim}
\usepackage{graphicx}
\usepackage{balance}
\usepackage{xcolor}
\usepackage[colorlinks,
            linkcolor=green,  
            urlcolor=blue,  
            citecolor=red,        
            ]{hyperref}
\usepackage{filecontents}
\def\BibTeX{{\rm B\kern-.05em{\sc i\kern-.025em b}\kern-.08em
    T\kern-.1667em\lower.7ex\hbox{E}\kern-.125emX}}

\PassOptionsToPackage{hyphens}{url}
\usepackage{cite}
\usepackage{mdwmath}
\usepackage{tabularx}
\usepackage{booktabs}
\usepackage{caption, threeparttable}
\usepackage[labelfont=bf]{caption} 
\usepackage{enumerate}
\usepackage{amssymb}
\usepackage{breakurl}
\usepackage{epstopdf}
\usepackage{mathrsfs}

\usepackage{algorithm} 

\newcommand{\Rmnum}[1]{\expandafter\@slowromancap\romannumeral #1@}
\renewcommand{\algorithmicrequire}{\textbf{Input:}}  
\renewcommand{\algorithmicensure}{\textbf{Output:}} 
\usepackage{multirow}
\usepackage[numbers,sort&compress]{natbib}

\makeatother
\begin{document}
\def \sn {\textsc{METANOIA}}
\title{\sn{}: A Lifelong Intrusion Detection and Investigation System for Mitigating Concept Drift}

\author{Jie Ying,
        Mengce Zheng,
        Jungan Chen,
        Ruoxi Chen,
        Zhongjie~Zhu*,
        Tiantian~Zhu*,

}

\markboth{IEEE Transactions on Dependable and Secure Computing, ~Vol.~xx, No.~yy, Month~2024}%
{\sn{}: A Lifelong Intrusion Detection and Investigation System for Mitigating Concept Drift}

\maketitle

\begin{abstract}
As Advanced Persistent Threat (APT) complexity increases, provenance data is increasingly used for detection. Anomaly-based systems are gaining attention due to their attack-knowledge-agnostic nature and ability to counter zero-day vulnerabilities. However, traditional detection paradigms, which train on offline, limited-size data, often overlook concept drift - unpredictable changes in streaming data distribution over time. This leads to high false positive rates. We propose incremental learning as a new paradigm to mitigate this issue.
\par
However, we identify \textbf{FOUR CHALLENGES} while integrating incremental learning as a new paradigm. First, the long-running incremental system must combat catastrophic forgetting (\textbf{C1}) and avoid learning malicious behaviors (\textbf{C2}). Then, the system needs to achieve precise alerts (\textbf{C3}) and reconstruct attack scenarios (\textbf{C4}). We present \sn{}, the first lifelong detection system that mitigates the high false positives due to concept drift. It connects pseudo edges to combat catastrophic forgetting, transfers suspicious states to avoid learning malicious behaviors, filters nodes at the path-level to achieve precise alerts, and constructs mini-graphs to reconstruct attack scenarios. Using state-of-the-art benchmarks, we demonstrate that \sn{} improves precision performance at the window-level, graph-level, and node-level by 30\%, 54\%, and 29\%, respectively, compared to previous approaches.
\end{abstract}


\section{INTRODUCTION}
\label{sec:1_intro}
\par
\IEEEPARstart{I}{n} comparison to raw audit logs, provenance graphs offer a novel perspective for identifying threats and investigating traces, facilitating the differentiation between prolonged normal behavior and malicious activities~\cite{han2020unicorn, nodoze, inam2022sok}. Presently, there is a growing trend in leveraging the rich contextual information provided by provenance graphs to perform host-based intrusion detection, referred to as \textit{provenance-based intrusion detection systems} (PIDSes)~\cite{hossain2017sleuth, morse, zhu2023aptshield, holmes, rapsheet, xiong2020conan, streamspot, FRAPpuccino, han2020unicorn, shadewatcher, threatrace, prographer, kairos, xie2018pagoda, SIGL, wang2020provdetector}.
\par
Among them, anomaly-based work defines a model of benign system behavior based on historical log data~\cite{streamspot, FRAPpuccino, han2020unicorn, shadewatcher, threatrace, prographer, kairos, xie2018pagoda, SIGL, wang2020provdetector}, triggering an alert when the execution significantly deviates from this model. Since it is attack-knowledge-agnostic, counteracts zero-day vulnerabilities, and possesses meticulous detection capabilities, many researchers focus on anomaly-based detection works.  Compared to the high latency and expenses associated with offline systems~\cite{shadewatcher, wang2020provdetector, SIGL, xie2018pagoda}, PIDSes capable of processing dynamic evolving provenance graphs in a streaming fashion~\cite{han2020unicorn, rapsheet, prographer, kairos, threatrace} are better suited for the real-time and cost-effective requirements of current host-based detection. 
\par
However, most current anomaly-based PIDSes seem to evade a critical issue: \textbf{concept drift}. Concept drift describes unforeseeable changes in the underlying distribution of streaming data over time~\cite{tsymbal2004conceptdrift, gama2014conceptdriftsurvey, lu2018conceptdriftlearning}. In this work, concept drift is represented as a variation in the benign behavior of a user on the host over time, e.g., switching work environments. This variation can lead to a decline in model performance since the features and patterns learned during training are no longer applicable during testing. In practical scenarios, this common phenomenon corresponds to changes in user behavior at the endpoint, such as switching to different production environments or deploying new software. This change (i.e., \textbf{concept drift}) can result in an unacceptable false-positive rate for existing PIDSes~\cite{inam2022sok, dong2023we, zipperle2022provenance}.
\par
We observe that current PIDSes adhere to the \textit{traditional anomaly detection paradigm}: training on offline-collected, limited-size data to identify anomalies (behaviors different from the training data) on test data. However, this paradigm does not work well as host behaviors exhibit infinite states and unknown variability. This uncertainty can lead to significant differences in the distribution of training and test data, i.e., \textbf{concept drift}. This situation causes the current anomaly-based PIDSes to have many false positives because they do not realize that the decision boundary at the current moment is different from the training time. Limitlessly expanding training data or strictly constraining user behaviors are two approaches to combat concept drift within this traditional paradigm, but these are not feasible.
\par
To feasibly mitigate concept drift (\textbf{Goal}), we introduce incremental learning~\cite{van2022threeincremental, he2011incremental} as a new anomaly detection paradigm (\textbf{Baseline}). Incremental learning refers to the model's ability to continuously assimilate dynamic data, enabling it to evolve and adapt within an ever-changing environment. Similarly, the system-level streaming logs are dynamic, with their distribution changing over time and continually introducing new entities and interactions. The correspondence between the continuous learning capability of incremental learning and the distribution variability of streaming logs prompts us to \textit{consider adopting incremental learning as a new anomaly detection paradigm to mitigate concept drift}.
\par
However, directly integrating incremental learning with anomaly-based PIDSes would diminish their intrusion detection capabilities. After thorough research analysis, we identified \textbf{FOUR CHALLENGES} that need to be overcome to integrate incremental learning to mitigate concept drift. First, long-running incremental PIDSes must combat catastrophic forgetting (\textbf{C1})~\cite{febrinanto2023lifelongsurvey}. Furthermore, the PIDSes should endeavor to avoid high false negatives introduced by learning malicious behaviors (\textbf{C2})~\cite{faber2023lifelongforanomalydetection}. Lastly, the PIDSes must achieve precise alerts (\textbf{C3}) and reconstruct attack scenario (\textbf{C4})~\cite{kairos, prographer, inam2022sok}. It is important to emphasize that these interconnected challenges were progressively uncovered as we attempted to mitigate concept drift. Subsequently, we will elaborate on these challenges.
\par
\textbf{C1 Combat Catastrophic Forgetting.} Some existing studies~\cite{shadewatcher, kairos, prographer} have adopted the concept of incremental learning, retraining models through \textit{oracles} (human guidance). However, this approach is plagued by \textit{catastrophic forgetting, a phenomenon where models, upon learning new tasks, gradually discard or forget how to perform previous tasks.} We argue that anomaly-based PIDSes should be capable of combating catastrophic forgetting, retaining memory of old tasks to reduce redundant training costs and minimize disruptive false alarms.
\par
\textbf{C2 Avoid Discrimination Paradox.} The Rehearsal Approach is a widely acknowledged and effective strategy for mitigating catastrophic forgetting~\cite{febrinanto2023lifelongsurvey}. It fundamentally aids models in retaining the memory of prior learning by revisiting old knowledge. 
Nonetheless, a specific challenge arises when directly implementing the Rehearsal Approach within endpoint detection: a \textit{discrimination paradox}. In detail, if the detection system can determine the maliciousness of data selected for replay, subsequent steps (detecting malicious behavior) become redundant. If it fails in this determination, the replay of malicious data might degrade detection performance. Consequently, PIDSes must navigate away from the discrimination paradox.
\par
\textbf{C3 Achieve Precise Alerts.} During prolonged deployment, PIDS often generates a significant number of false alarms due to the change in execution environment, the introduction of new applications, and the occurrence of low-frequency operations. It is imperative to recognize that these alarms typically reflect anomalous, yet not necessarily malicious, behaviors within the system.  Therefore, we contend that PIDSes should strive to minimize such false positives caused by concept drift, ensuring precise alerts and reducing the additional workload on security practitioners.
\par
\textbf{C4 Reconstruct Attack Scenario.} Provenance data can assist in investigating the attack chains leading to intrusions and the potential damage caused by intrusions to the system~\cite{nodoze, sleuth, king2003backtracking}. However, manually investigating the entire graph is not feasible due to the substantial size of a typical provenance graph and its rapid expansion rate over time~\cite{pasquier2017practical}. Practical PIDSes need to reconstruct attack scenarios through dependencies among kernel objects, significantly reducing manual work and enabling system administrators to quickly understand intrusions and timely design responses.
\par
To this end, we introduce \sn{}, the first lifelong anomaly detection and investigation system that utilizes incremental learning as a new paradigm to address the high false positive problem caused by concept drift.  Where lifelong refers to \sn{}'s ability to continuously process streaming data to detect anomalies over a long period. \sn{} connects pseudo edges to combat catastrophic forgetting, transfers suspicious states to avoid the discrimination paradox, filters at the path-level to achieve precise alerts, and constructs mini-graphs to reconstruct attack scenarios. In summary, this paper makes the following contributions:

\begin{itemize}
\item{We conduct an in-depth analysis of the concept drift problem in current anomaly-based PIDSes, present shortcomings of the traditional anomaly detection paradigm, and finally propose a novel anomaly detection paradigm.}

\item{We introduce \sn{}, a lifelong intrusion detection and investigation system for mitigating concept drift. \sn{} identifies, analyzes, and addresses four challenges inherent in adopting a new anomaly detection paradigm. We believe that \sn{} effectively minimizes false positives stemming from concept drift. To our best knowledge, \sn{} is the first system to address concept drift through the adoption of a novel anomaly detection paradigm.}

\item{We evaluate \sn{} on publicly available benchmark datasets from DARPA~\cite{darpa-tc} that simulate APT campaigns, as well as datasets that allow us to fairly compare \sn{} with state-of-the-art PIDSes, UNICORN~\cite{han2020unicorn}, KAIROS~\cite{kairos}, PROGRAPHER~\cite{prographer} and ThreaTrace~\cite{threatrace}. Experimental results show that the precision performance of \sn{} on window-level, graph-level, and node-level outperforms existing approaches by 30\%, 54\%, and 29\% respectively.}

\end{itemize} 

\section{MOTIVATION}
\label{sec:2_motivation}

To motivate this work, we first describe a specific scenario for PIDSes and then present prior approaches with their limitations.

\begin{figure}[h!t]
\centering
\includegraphics[width=\linewidth]{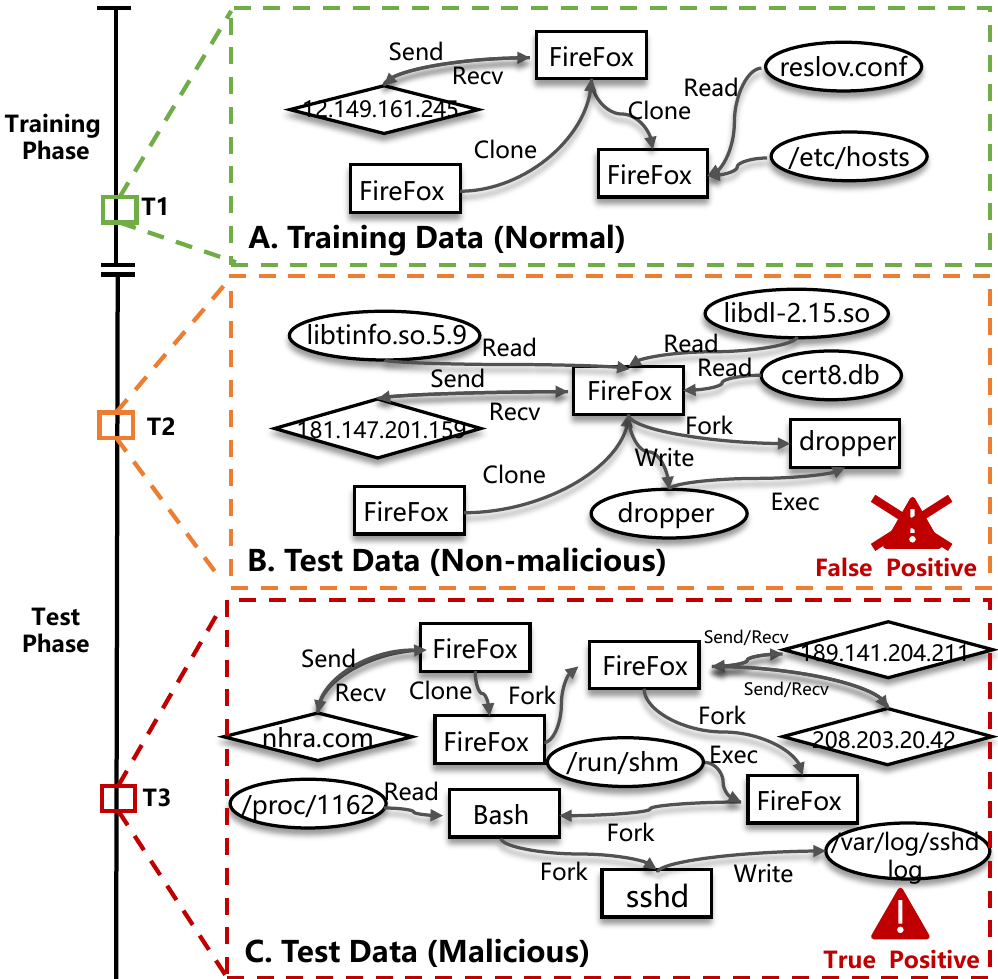}
\caption{Motivating Example. 
}
\label{fig:motivation_example}
\vspace{-0.3cm}
\end{figure}

\subsection{Scenario} As shown in Figure~\ref{fig:motivation_example}, the entire lifecycle of PIDSes can be divided into two stages: the training phase (including Behavior A) and the testing phase (including Behavior B and C). During the training stage, PIDSes model itself based on collected normal behaviors and establishes specific decision boundaries. And according to these boundaries, PIDSes identify any anomalies deviating from them and raise an alert in the testing phase.
\par
Behavior A and Behavior B are both non-malicious behaviors of the system, and Behavior C is a large-scale APT campaign simulated by DARPA~\cite{darpa5}. Since Behavior B and Behavior C are contextually and semantically different from Behavior A, PIDSes recognize them as anomalies and repeat threat alerts. 
\par
\textbf{Behavior A (at time T1)}: The Firefox process first clones itself, then connects to \texttt{12.149.161.245}, and subsequently reads the \texttt{resolv.conf} and \texttt{/etc/hosts} files. 
\par
\textbf{Behavior B (at time T2)}: The Firefox process first clones itself, then connects to \texttt{181.147.201.159}, and proceeds to read the \texttt{cert8.db} and \texttt{libdl-2.15.so} files. Finally, it writes and executes the \texttt{dropper}. 
\par
\textbf{Behavior C (at time T3)}: The attack begins with the compromise of a once-reliable website (\texttt{nhra.com}), where the attacker lies in wait for an individual using a vulnerable Firefox browser to establish a link with a malicious server. When the individual accesses the site, the attacker proceeds to run a Drakon exploit in the memory of Firefox. Consequently, Firefox is forced to establish a connection with an attacker-controlled server at the IP addresses: \texttt{189.141.204.211} and \texttt{208.203.20.42}. By exploiting a driver previously installed, named “\texttt{/run/shm}”, the attacker escalates privileges on the Firefox process and achieves root access to create a new process \texttt{Bash} with access to process list. In the final stage, the attacker employs the Inject2 Process injection method to introduce shellcode into the \texttt{sshd}, resulting in the creation of a file named \texttt{sshdlog} on the disk. This action paves the way for the attacker to reach out to additional harmful web servers causing data exfiltration. 

\subsection{Prior Approaches \& Limitations}
\label{sec:2.1_prior_approache}
Traditional supervised methods, such as ATLAS~\cite{alsaheel2021atlas} and APT-KGL~\cite{chen2022aptkgl}, require the use of attack samples as training data for manipulation. However, in practice, acquiring such data is extremely challenging. To address this issue, recent PIDSes have adopted unsupervised or semi-supervised detection pipelines. These pipelines model benign data and identify abnormal behaviors as indications of malicious activities. Among state-of-the-art PIDSes, Unicorn~\cite{han2020unicorn} and StreamSpot~\cite{streamspot} provide graph-level detection (representing the current state of the entire system), ThreaTrace~\cite{threatrace} and MAGIC~\cite{jia2023magic} offer entity-level detection, while KAIROS~\cite{kairos} and ShadeWatcher~\cite{shadewatcher} focus on event-level detection.
\par
We observe that current PIDSes follow the same paradigm, as shown in Figure~\ref{fig:motivation_example}: training on offline-collected, limited-size data to identify anomalies (behaviors different from the training data) on test data. Unfortunately, this paradigm are not suitable for endpoint anomaly detection, as host behaviors exhibit infinite states and unknown variability. This uncertainty can lead to significant differences in the distribution of training and test data, i.e., \textbf{concept drift}. It is the situation that brings about a large number of false positives for current PIDSes because they fail to realize that the decision boundary has changed after the training stage.
\par
As shown in Figure 2, PIDSes only modeled the Behavior A  of Firefox during the training phase. In the testing phase, Firefox exhibited Behavior B, which was deemed anomalous by PIDSes (as it crossed the decision boundary). We believe this phenomenon is quite common, such as when users deploy new software or change their production environment. PIDSes triggered an alert (false positive), even though the Behavior B of the Firefox process at time T2 was non-malicious. This discrepancy arises from the fact that existing PIDSes decision boundaries did not adapt to changes in data distribution.
\par
To sum up, we believe that the traditional paradigm is limited by high false positive rates in anomaly detection scenarios with concept drift. Therefore, \sn{} employs a novel anomaly detection paradigm to mitigate false positives induced by concept drift.

\section{Threat Model}
\label{sec:4_threatmodel}
We make the following assumptions about the target system. The Trusted Computing Base (TCB) of the system consists of the operating system and audit framework. We assume these components are operating correctly and cannot be manipulated by attackers. We also assume that attackers cannot directly modify the contents of audit logs, which can be achieved through tamper-evident logging techniques~\cite{karande2017sgx, paccagnella2020custos, paccagnella2020logging, yagemann2021validating} beyond the scope of this paper. Finally, we do not consider hardware-level, side-channel, or convert-channel attacks, since kernel-level audit systems do not explicitly capture their behaviors.
\par
It is worth emphasizing that we do not assume the absence of concept drift, which distinguishes our approach from the majority of studies~\cite{lee2013loggc, ma2016protracer, ma2017mpi, kwon2018mci, kairos, shadewatcher, threatrace, prographer}. These studies typically assume that system behaviors remain static over time, i.e., no concept drift exists. However, host behaviors evolve in real-world scenarios with changing user activities and requirements. Therefore, we refrain from making such unrealistic assumptions and instead, \sn{} can serve as an intrusion detection and investigation system for concept drift mitigation using system-level provenance.

\section{Framework}
\label{sec:4_overview}
\begin{figure*}[h!t]
\centering
\includegraphics[width=\linewidth]{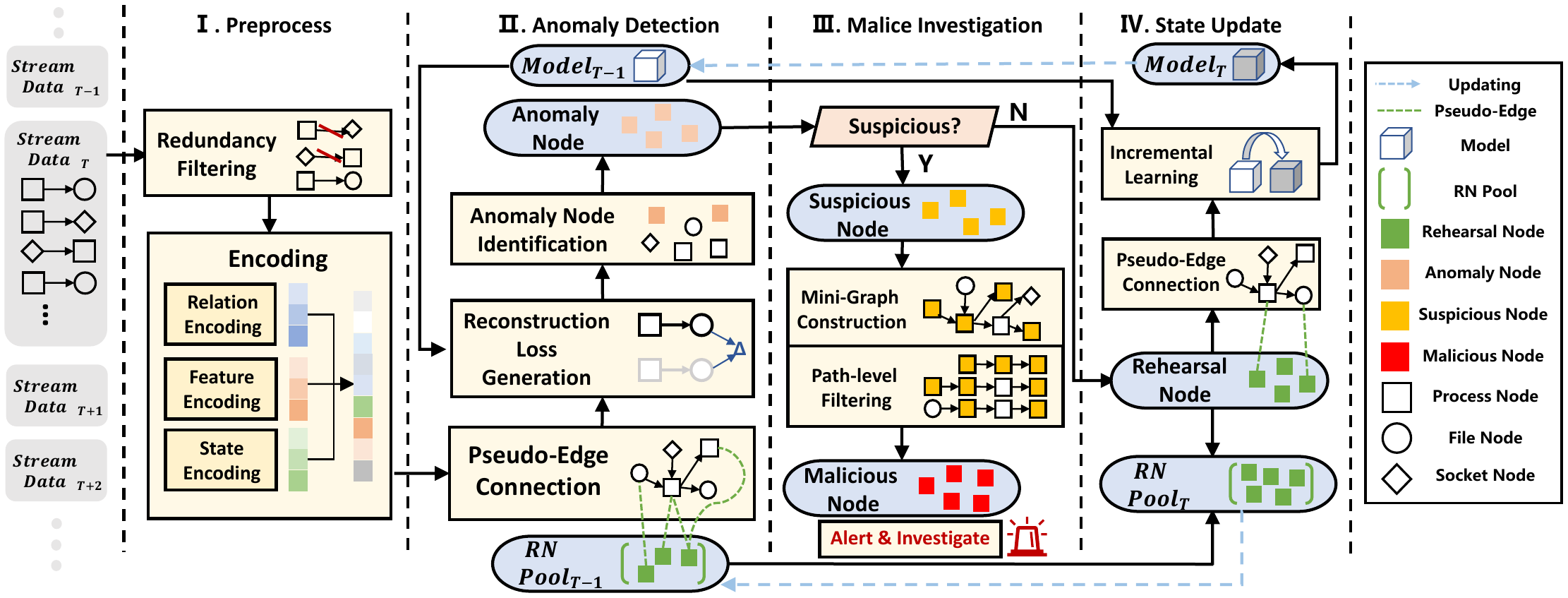}
\caption{ Overview of \sn{}'s architecture. 
}
\label{fig:approach}
\vspace{-0.8cm}
\end{figure*}
In this section, we first introduce four critical node types involved in \sn{} and then provide an overview of \sn{}.

\subsection{Critical Node Types}
\sn{} considers that process nodes to exhibit the richest behavioral characteristics and hold the highest identification status during intrusion detection. Additionally, given the long-term nature of APT attack detection, the four types of nodes defined by \sn{} are all process nodes.
\par
\textbf{Anomalous Nodes (AN)} refer to nodes that exhibit new behavior compared to the past. \sn{} believes that anomalous nodes typically consist of two distinctive behaviors: (1) nodes exhibiting behavior inconsistent with their past experiences, such as a user who predominantly uses a browser for work suddenly engaging in leisure activities, and (2) newly initiated nodes, such as users installing new applications to accomplish production goals. AN occur in large numbers when the host switches working environments or runs new applications. 
\par
\textbf{Suspicious Nodes (SN)} refer to anomalous nodes with suspicious state, while \textbf{Rehearsal Nodes (RN)} refer to anomalous nodes without suspicious state. In other words, the union of suspicious and rehearsal nodes constitutes anomalous nodes. \sn{} believes that rehearsal nodes serve to help the model memorize scenarios that are anomalous rather than malicious, thereby reducing false positives caused by concept drift. Consequently, \sn{} selects anomalous nodes without suspicious states as rehearsal nodes.


\par
\textbf{Malicious Nodes (MN)} refer to nodes associated with attacks. \sn{} obtains malicious nodes by filtering suspicious nodes, which means that malicious nodes must be a subset of suspicious nodes. \sn{} initiates an alert when a malicious node is recognized and connects these nodes to provide a concise attack scenario graph.

\subsection{Overview}
\label{sec:4.2_overview}
\sn{} is an unsupervised online PIDS that effectively detects intrusion while mitigating the high false positives caused by concept drift. \sn{} adopts a novel anomaly detection paradigm by incorporating incremental learning, i.e., learning changing data distributions and updating model parameters as time windows on streaming data.  
\par
The basic architecture of \sn{} is shown in Figure~\ref{fig:approach}, which can be divided into four phases: (\uppercase\expandafter{\romannumeral1}) Preprocess, (\uppercase\expandafter{\romannumeral2}) Anomaly Detection, (\uppercase\expandafter{\romannumeral3}) Malice Investigation, and (\uppercase\expandafter{\romannumeral4}) State Update. It is important to note that \sn{} iteratively loops the above four phases as the time window progresses. For clarity, we will only describe the overview of \sn{} within a specific time window (denoted as time window \textit{T} in Figure~\ref{fig:approach}). Details of system design for each phase are given in Section~\ref{sec:5_sysdesign}.
\par
\textbf{Preprocess (§~\ref{sec:5.1_preprocess}).} \sn{} processes streaming data at the level of time windows, which is very common in existing work~\cite{kairos, prographer, li2023nodlink}. \sn{} will perform redundancy filtering of events in the window and encode the events in terms of relation, feature, and state to achieve embedding. 
\par
\textbf{Anomaly Detection (§~\ref{sec:5.2anomaly_detection}).} When a new edge emerges, \sn{} connects pseudo-edges based on the reachability between rehearsal nodes in the RN Pool and the subject and object of the new edge. Next, \sn{} generates embedding containing the information on rehearsal nodes, the neighborhood structure around the edge, and the states of nodes in the neighborhood. Then, \sn{} reconstructs the edge based on this embedding and generates reconstruction loss with the original edge. Finally, \sn{} identifies AN in this time window by utilizing the reconstruction losses. Finally, these reconstruction losses are utilized by \sn{} to identify the AN in this window.
\par
\textbf{Malice Investigation (§~\ref{sec:5.3_maliciousInvestigation}).} \sn{} selects AN with suspicious state (i.e., the results of encoded states in preprocess) as suspicious nodes. Next, \sn{} constructs a mini-graph of suspicious nodes for the current window based on the Steiner Tree algorithm~\cite{hwang1992steiner, imase1991dynamicsteiner} and merges it with the global mini-graph, which is called mini-graph construction. Afterward, \sn{} scores and filters suspicious nodes at the path level to reduce false positives caused by concept drift as much as possible. Finally, \sn{} treats the retained suspicious nodes as malicious nodes, triggers an alert, and reconstructs the attack scenario graph associated with these malicious nodes.
\par
\textbf{State Update (§~\ref{sec:5.4_stateUpdate}).} 
\sn{} merges rehearsal nodes into the previous $RN\ Pool_{T-1}$ to get a new $RN\ Pool_{T}$ for pseudo-edge connections. Following that, \sn{} employs the current $Stream\ Data_{T}$ for training the previous $Model_{T-1}$ in order to obtain a $Model_{T}$ with revised parameters. $Model_{T}$ is then used to generate reconstruction losses (Phase \uppercase\expandafter{\romannumeral1})  and support incremental learning (Phase \uppercase\expandafter{\romannumeral4}) for the next $Window_{T+1}$.

\section{SYSTEM DESIGN}
\label{sec:5_sysdesign}
We will explain the implementation details of each phase of \sn{} in this section.

\subsection{Preprocess}
\label{sec:5.1_preprocess}
\sn{} constructs a whole-system provenance graph from streaming data collected by logging infrastructures, such as Windows ETW~\cite{etw}, Linux Audit~\cite{audit}, and CamFlow~\cite{pasquier2017camflow}. \sn{} considers three types of kernel objects and eight types of interactions. \sn{} transforms each event into a directed, time-stamped edge, in which the source node represents the event's subject and the destination node represents the object being acted upon. For any event $ e \in E$, \sn{} represents it as a quintuple $\langle{U_{s}, U_{o}, O, T_{i}}\rangle$. $U_s$ and $U_o$ are unique identifiers for the subject and object of $e$, respectively. $O$ denotes the type of $e$, and $T_i$ denotes the time when $e$ occurred. Table~\ref{table:EntityEventType} shows the events between subjects and objects and the node attributes we consider. As shown in Figure~\ref{fig:approach} (Phase \uppercase\expandafter{\romannumeral1}), \sn{} will perform redundancy filtering and encoding in preprocessing, as described in more detail below.

\subsubsection{Redundancy Filtering} 
\label{5.1.1_redundancy_filtering}
In order to reduce redundant audit records when detecting and investigating cyber threats, we employed several redundancy filtering techniques from previous works in processing streaming data. Firstly, we implemented the Causality Preserved Reduction (CPR)~\cite{xu2016cpr} to aggregate audit records with the same impact scope. For instance, in Figure~\ref{fig:motivation_example}.C, \texttt{FireFox} receives network packets from \texttt{nhra.com} up to 198 times in a single data transmission operation. This is because the operating system typically distributes data among multiple system calls proportionally when performing tasks such as file read/write events. We believe that merging such redundant events does not alter the transmission of information. Another type of redundancy~\cite{tang2018nodemerge} that needs filtering involves events between processes and read-only libraries during initialization. To address this, we established a whitelist of trusted libraries (e.g., "libc.so", "libm.so") and removed them without compromising the correctness of causal analysis.

\subsubsection{Encoding}
\label{sec:5.1.2_encoding}
\sn{} primarily encodes events from three perspectives: relation, feature, and state. Firstly, it is necessary to clarify that relation refers to the type of event, feature denotes the attribute of node (e.g., process/file name), and state indicates the semantic states of node.
\par
For any event $ e \in E$, \sn{} conducts feature encoding and state encoding for $U_s$ and $U_o$ separately (FE($U_s$), SE($U_s$), FE($U_o$), SE($U_o$)), and encodes $O$ as its type (RE($O$)). Subsequently, these five encoded results will be concatenated to serve as the final encoding of this event for future use in Section~\ref{sec:5.2anomaly_detection}.
\par
Relation Encoding: \sn{} utilizes one-hot encoding to embed events, similar to most PIDSes~\cite{prographer, kairos, shadewatcher, SIGL}. We consider this one-hot encoding approach to be sufficiently simple and effective in assisting the model to differentiate between different types of relations in a flat label space.
\par
Feature Encoding: \sn{} utilizes hierarchical feature hashing~\cite{zhang2020dynamic} to encode node features. Hierarchical feature hashing employs a hash function to map original features into a fixed-size hash space, thereby transforming high-dimensional features into low-dimensional ones. 
This encoding method is naturally suitable for file paths with hierarchical structures, making it more likely for file names with similar semantics to be mapped to proximate positions in the hash space. For example, the distance between \texttt{/usr/local/share} and \texttt{/usr/local/lib} in feature space will be less than the distance to \texttt{/usr/ports/shells}.
\par
State Encoding: \sn{} encodes the state of nodes from the perspective of information/control flow.  \sn{} defines the suspicious node (possibly related to attacks) using a suspicious state $SS$, a real number ranging from 0 to 1. The closer value to 1 the higher level of suspicion. \sn{} initializes processes and files as benign ($SS=0$). Given that \sn{} assumes attackers intrude from external sources by default, all sockets are initialized as suspicious ($SS=1$).
\par
In addition, \sn{} devises \textbf{Suspicious State Transfer Rules} to convey the states of streaming data, as illustrated in Table~\ref{tab:sus_semantic_rules}. For example, if a freshly started benign process like \texttt{nginx} ($SS=0$) reads a file with a suspicious state ($SS=1$), such as \texttt{/tmp/mal.sh}, then \texttt{nginx} will inherit the suspicious state. 
\par
Finally, due to the issue of dependency explosion, the overspreading of states may render the node states meaningless. Hence, \sn{} introduces a decay factor $\beta$ to ensure that the node states decay with the increase in propagation distance. The specific setting of the $\beta$ is discussed in detail in Section~\ref{sec:6_evaluation}.



\begin{table}[]
\centering
\caption{Entities, their attributes, and event types.}
\label{table:EntityEventType}
\begin{tabular}{|c|c|c|c|}
\hline
\textbf{Subject}         & \textbf{Object} & \textbf{Events}         & \textbf{Entitiy Attributes} \\ \hline
\multirow{3}{*}{Process} & Process         & Fork, Clone             & Process Name                \\ \cline{2-4} 
                         & File            & Read, Write, Mmap, Exec & File Name                   \\ \cline{2-4} 
                         & Socket          & Sendto, Recvfrom        & Src/Dst IP/Port             \\ \hline
\end{tabular}
\end{table}

\begin{table}[]
\centering
\caption{Suspicious State Transfer Rule. In the second column, S denotes the subject and O denotes the object. In the third column, $S.ss$ and $O.ss$ denote the subject's and object's suspicious state, and $\beta$ denotes the decay factor.}
\label{tab:sus_semantic_rules}
\begin{tabular}{l|l|l}
\hline
\textbf{Event} & \textbf{Direction} & \textbf{Suspicious Delivery} \\ \hline
Recvfrom       & $O\to S$               &   $\beta \cdot max(O.ss, S.ss) \Rightarrow S.ss$                  \\
Read/Mmap/Exec & $O\to S$               &  $\beta \cdot max(O.ss, S.ss) \Rightarrow S.ss$                  \\
Write          & $S\to O$               & $\beta \cdot max(O.ss, S.ss) \Rightarrow O.ss$                 \\
Fork/Clone     & $S\to O$               & $\beta \cdot max(O.ss, S.ss) \Rightarrow O.ss$                 \\ \hline
\end{tabular}
\vspace{-0.3cm}
\end{table}


\subsection{Anonmaly Detection}
\label{sec:5.2anomaly_detection}
\par
In this section, we sequentially describe the process of Phase \uppercase\expandafter{\romannumeral2} in Figure~\ref{fig:approach}, i.e., pseudo-edges connection, reconstruction loss generation, and anomalous node identification. In short, \sn{} connects pseudo-edges to introduce anomalous scenario information associated with rehearsal nodes, generates reconstruction loss to determine the abnormality level of the current event, and identifies anomalous nodes in preparation for Phase \uppercase\expandafter{\romannumeral3}.
\subsubsection{Pseudo-Edge Connection} 
\label{sec:5.2.1_pseudoEdgeConnection}
When \sn{} adopts the incremental learning paradigm to address concept drift, it encounters the issue of \textit{catastrophic forgetting}. This situation leads the model to forget the previously learned knowledge as it learns new one. This problem can result in high false positives and significant retraining costs for anomaly detection systems in long-term operation.
\par
Inspired by the classic network architecture ResNet~\cite{he2016resnet}, \sn{} attempts to solve the problem by connecting pseudo-edges for scene replay. There are two key insights of \sn{}: \textit{(1) Rehearsal nodes are anomalous yet benign, containing critical information about anomalous scenes}, and \textit{(2) Pseudo-edge connection assist the model in revisiting anomalous scenes and expanding the scope of recollection.} For any event $e$, \sn{} assesses whether a reachable relationship exists between its entity ($U_s$ or $U_o$) and the rehearsal nodes in the RN Pool. If so, \sn{} establishes a pseudo-edge between them to replay crucial scene information.  \sn{} considers that this method helps the model combat catastrophic forgetting through scene recall and prevent the inclusion of malicious scenes. It should be noted that pseudo-edges are only temporarily constructed during the generation of event embedding and do not affect the original components of the provenance graph.
\par
This subsection presents the principle and method of pseudo-edge construction. The details on how to utilize pseudo-edge information and select rehearsal nodes will be discussed in Sections~\ref{sec:5.2.2_reconstructionLossGeneration} and~\ref{sec:5.4_stateUpdate}, respectively.

\subsubsection{Reconstruction Loss Generation}
\label{sec:5.2.2_reconstructionLossGeneration}
We believe that only by considering all causally related events and their temporal order can one determine whether a node is anomalous. \sn{} employs Temporal Graph Networks (TGN)~\cite{rossi2020temporal} to reconstruct event types, and generate reconstruction loss (the difference between reconstructed types and actual types).

\par
For each node $i$ seen so far, \sn{} maintains a memory vector $s_i$ representing the node's history in a compressed format. For each event involving node $i$, a message is computed to update $i$'s memory. In the case of an interaction event $e_{ij}(t)$ between source node $i$ and target node $j$ at time $t$, the message $m_{i}(t)$ can be computed by:
{\setlength\abovedisplayskip{0.1cm}
\setlength\belowdisplayskip{0.1cm}
\begin{equation}
\begin{aligned}
&m_{i}(t)\  = \ msg_{s}(s_{i}(t^{-}), \ s_{j}(t^{-}), \  \bigtriangleup t, \  e_{ij}(t)) \\
\end{aligned}
\end{equation}}
where $s_{i}(t^{-})$ is the memory of node $i$ just before time $t$ and $msg_{s}$ is the concatenation operation. The $\bigtriangleup t$ refers to the difference between $t^{-}$ and $t$, which \sn{} uses to introduce temporal information. For interaction events involving two nodes $i$ and $j$, the memories of both nodes are updated after the event. \sn{} updates the node's memory for each event involving the node itself:
{\setlength\abovedisplayskip{0.1cm}
\setlength\belowdisplayskip{0.1cm}
\begin{equation}
\begin{aligned}
&s_{i}(t)\  = \ mem (m_{i}(t), s_{i}(t^{-}))\\
\end{aligned}
\end{equation}}
where $mem$ is a learnable function and \sn{} uses the GRU~\cite{cho2014gru}. Then, \sn{} generates the final embedding $z_{ij}(t)$ for edge $e_{t}$ at any given time $t$:
{\setlength\abovedisplayskip{0.1cm}
\setlength\belowdisplayskip{0.1cm}
\begin{equation}
\begin{aligned}
&z_{ij}(t)=\sum_{j\in N(i)}h(s_{i}(t), \ s_{j}(t), \ e_{ij}(t), \ v_{i}(t), \ v_{j}(t)) 
\\
\end{aligned}
\end{equation}}
where $h$ is a learnable function, and \sn{} chooses an attention-based Graph Neural Network (GNN)~\cite{shi2020masked}. $N(i)$ refers to all neighboring nodes of node $i$. $e_{ij}$ refers to the encoding of event $e$ by \sn, which is described in detail in Section~\ref{sec:5.1.2_encoding}.  $v_{i}(t)$ refers to the temporal feature of node $i$ at time $t$, which is generated after \sn{} processes the timestamp using a simple linear mapping. 
In summary, \sn{} embeds the events from node historical information ($m_{i}(t)$), event encoding ($e_{ij}(t)$), and neighborhood structures ($N(i)$).  
\par

\par
Finally, \sn{} employs a Multi-Layer Perceptron (MLP) to transform the embedding vector $z_{ij}(t)$ into vector $P(e_t)$. Then \sn{} utilizes cross-entropy to compute the loss between $P(e_t)$ and the observed event type (in one-hot form) as the Reconstruction Loss (RL).

\subsubsection{Anomaly Node Identification}
\label{sec:5.2.3_anomalyNodeIndentification}
\sn{} identifies anomalous nodes based on the reconstruction loss. Specifically, \sn{} calculates a threshold $\sigma_{t}$ in each time window $t$, which is the mean of all reconstruction losses within that window plus two standard deviations. \sn{} will consider the relevant subjects and objects of all events with reconstruction losses exceeding this threshold $\sigma_{t}$ to be anomalous nodes.


\subsection{Malice Investigation}
\label{sec:5.3_maliciousInvestigation}
Unlike other existing approaches~\cite{prographer,shadewatcher,han2020unicorn}, \sn{} does not directly equate anomaly with maliciousness and issue an alert upon detecting AN but enters a secondary investigation. As shown in Figure~\ref{fig:approach} (Phase \uppercase\expandafter{\romannumeral3}), \sn{} firstly combines the information of state encoding (Section~\ref{sec:5.1_preprocess}) to select AN (the output of Phase \uppercase\expandafter{\romannumeral2}) with suspicious states higher than the suspicion threshold $\gamma$ as suspicious nodes. In other words, suspicious nodes are those nodes that simultaneously exhibit suspicious semantics and anomalous behavior. Next, \sn{} obtains the final malicious nodes through mini-graph construction and path-level filtering oriented towards suspicious nodes. The detailed design is as follows:
\subsubsection{Mini-Graph Construction.} Typically, there are multiple suspicious nodes in each time window, and \sn{} posits that these suspicious nodes are not independent of each other. The key insight behind mini-graph construction: \textit{there are the most anomalous and shortest reachable relationships among suspicious nodes.} Based on this insight, \sn{} will construct a mini-tree $MT_{t}$ for all suspicious nodes within the current window $t$ by the Steiner Tree algorithm~\cite{hwang1992steiner, imase1991dynamicsteiner} combined with reconstruction loss. The Steiner Tree algorithm's primary objective is to locate a tree that encompasses a designated set of specific nodes (i.e., suspicious nodes) in the graph (i.e., provenance graph $G_{t}$ in time window $t$) while minimizing the total weight of the tree as much as possible. Due to the interchangeability between maximum and minimum, \sn{} can utilize this algorithm to find the mini-tree $MT_t$, which contains all suspicious nodes within the time window $t$ and has the largest reconstruction loss sum. For a new $MT_t$, \sn{} either merges it into the existing mini-graph $MG$ or creates a new mini-graph containing only $MT_t$.  That is, $MT_t$ is merged into the existing $MG$: 
{\setlength\abovedisplayskip{0.1cm}
\setlength\belowdisplayskip{0.1cm}
\begin{equation}
\begin{aligned}
&MG = MG\cup MT_{t}, if\ MG\cap MT_{T} \ne \emptyset  \\
\end{aligned}
\end{equation}}
If the new $MT_t$ has no intersection with any $MG$, then the $MT_t$ is preserved as a new $MG$.
\subsubsection{Path-level Filtering.} 
\label{sec:5.3.2_pathlevel_filtering}
The goal of \sn{} is to achieve precise intrusion detection while mitigating the high false positives problem caused by concept drift. Through observation, \sn{} has identified two typical false positives: (1) New process initiation. Since \sn{} is an anomaly-based detection model, it inevitably identifies unseen processes as suspicious nodes, leading to false alarms. (2) Long-running suspicious processes. Legitimate processes that frequently interact with the external environment may carry a suspicious label for an extended period. \sn{} does not consider them as rehearsal nodes, thus unable to memorize their behavioral characteristics over the long term, resulting in false alarms. 
\par
Based on the observation, \sn{} designs a path-level filtering algorithm based on feature scoring to eliminate these two types of false positives, as shown in Algorithm~\ref{algo:path_level_filtering}. Firstly, for any suspicious node $sn_{i}$ under time window $t$, \sn{} employs random walks on its corresponding MG to generate multiple paths (lines 1-3). Each path is comprised of traversed nodes and events, including information such as node indexes, event reconstruction losses, and elapsed time windows. Subsequently, \sn{} scores all paths from four perspectives: volatility, periodicity, heterogeneity, and persistence (lines 7-11). \sn{} considers the path with high volatility (significant reconstruction loss discrepancies), weak periodicity (erratic node behavior), strong heterogeneity (a multitude of non-repeating nodes in the path), and high persistence (long duration) to be more likely traces left by a malicious node. Specifically, \sn{} employs Standard Deviation to measure volatility (line 8), Discrete Fourier Transform to calculate periodicity (line 9), the de-duplicated set of node names divided by the path length as heterogeneity (line 10), and path length as persistence (line 11). As these four features are not on the same scale, \sn{} normalizes each separately (line 12). Finally, \sn{} generates the final path score based on these four features and identifies the suspicious node $sn_{i}$ with the presence of a path score above the threshold $\delta$ (malicious path) as a malicious node (lines 13-15).

\begin{algorithm}[!h]
    \caption{PATH-LEVEL FILTERING}
    \label{algo:path_level_filtering}
    \renewcommand{\algorithmicrequire}{\textbf{Input:}}
    \renewcommand{\algorithmicensure}{\textbf{Output:}}
    \begin{algorithmic}[1]
        \REQUIRE Suspicious Nodes $SN$ at Time Window $t$
            \\   Mini-Graph $MG$ 
            \\   Path-level Scoring Threshold $\delta$
        \ENSURE Malicious Nodes $MN$ 
        \FOR{each $sn_{i}$ in $SN$}
            \IF{$sn_{i}\in MG$}
                \STATE PATHS = RandomWalk($sn_{i}$, MG) 
            \ELSE
                \STATE CONTINUE
            \ENDIF
            \FOR{each $path$ in PATHS}
                \STATE Volatility \ \ \ \ = StandardDeviation($path$)
                \STATE Periodicity \ \ \ = 1 / DiscreteFourierTransform($path$)
                \STATE Heterogeneity = Set(NodeName)/Len($path$)
                \STATE Persistence \ \ \ = Len($path$)
                \STATE Normalize(Volatility, Periodicity, Heterogeneity, Persistence)
                \STATE PathScore \ \ \ = Sum(Volatility, Periodicity, Heterogeneity, Persistence)
                \IF{PathScore $>$ $\delta$}
                    \STATE $MN[sn_{i}]$.append($path$)
                \ENDIF
            \ENDFOR
        \ENDFOR
        \RETURN $MN$
    \end{algorithmic}
\end{algorithm}

\par
\subsubsection{Alert \& Investigate}
After identifying a malicious node within a window, \sn{} reconstructs the corresponding attack scenario of that malicious node. Considering the persistence of APT attacks, for those suspicious nodes that are not identified as malicious, \sn{} considers the possibility that they may contain latent malicious semantics (for example, if attackers are still in the discovery phase). To avoid false negatives resulting from learning malicious behavior, \sn{} discards these nodes instead of placing them into the RN Pool.
\par
Typically, attacks span multiple windows, requiring \sn{} to identify correlations between different windows to reconstruct the complete attack scenario. To begin with, \sn{} designates the current window as a malicious window. Next, it traverses preceding windows iteratively to determine if the malicious paths in the preceding windows intersect with those in the malicious window. If so, the preceding windows are also considered malicious. Then, \sn{} merges all malicious paths along with suspicious events (exceeding the reconstruction loss threshold $\sigma$ in Section~\ref{sec:5.2.2_reconstructionLossGeneration}) related to the malicious paths as the reconstructed attack scenario. Lastly, \sn{} triggers an alert and returns the attack scenario reconstructed with the malicious node.

\subsection{State Update}
\label{sec:5.4_stateUpdate}
\sn{} integrates incremental learning as a novel paradigm for anomaly detection and therefore requires state update over time, as shown in Figure~\ref{fig:approach} (Phase \uppercase\expandafter{\romannumeral4}). Initially, \sn{} identifies anomalous nodes with suspicion below the threshold $\gamma$ (Section~\ref{sec:5.3_maliciousInvestigation}) as rehearsal nodes, merging them with the old $RN\ Pool_{t-1}$ to form the new $RN\ Pool_{t}$. Subsequently, \sn{} follows the procedure in Section~\ref{sec:5.2.1_pseudoEdgeConnection} to utilize the $RN\ Pool_{t}$ for pseudo-edge connection. Finally, \sn{} trains the old $Model_{t-1}$ and updates it to the new model $Model_{t}$ using streaming data from time window $t$.
\par
If no rehearsal nodes are detected, it indicates that \sn{} ($Model_{T-1}$) is familiar with the data from the time window $t$ or that all anomalous behaviors originate from suspicious nodes. And, \sn{} updates $Model_{T-1}$ only after detecting rehearsal nodes.

\section{Evaluation}
\label{sec:6_evaluation} 
\sn{} is an intrusion detection system designed to address the high false positive problem caused by concept drift. Therefore, we focus on addressing the following questions:
\begin{itemize}
    \item \textbf{RQ 1}: Can \sn{} achieve better detection performance than SOTA methods?
    \item \textbf{RQ 2}: How important are the components we designed for mitigating concept drift?
    \item \textbf{RQ 3}: Can \sn{} effectively reconstruct attack scenarios?
    \item \textbf{RQ 4}: How do hyperparameters affect the performance of \sn{}?
    \item \textbf{RQ 5}: Is \sn{} fast enough to perform realtime detection and investigation?
\end{itemize}

\subsection{Experiment Protocol}
We implemented a \sn{} prototype in Python. We utilized PyG~\cite{pyg} to generate reconstruction losses, Networkx~\cite{networkx} to construct the minimum graph, and GraphViz~\cite{graphviz} to visualize the reconstructed attack scenario graph. \sn{} leverages the agents to collect data from monitored hosts. The collected data have the same format as other detection systems~\cite{kairos, prographer, shadewatcher, han2020unicorn}.
\par
We selected three representative anomaly-based APT online detection systems, namely UNICORN~\cite{han2020unicorn}, PROGRAPHER~\cite{prographer}, and KAIROS~\cite{kairos}, as baselines for comparison. And we will compare the window-level detection accuracy with these three and node-level detection accuracy with the latter two. It is noteworthy that PROGRAPHER and KAIROS did not directly evaluate their node-level detection capabilities in the original paper. However, we observed that both reconstruct attack scenarios, aligning with \sn{}'s functionality. Therefore, we considered the nodes involved in reconstructed attack scenarios as the basis for evaluating node-level detection accuracy. We directly used the source code released by UNICORN and KAIROS and employed their default parameters to avoid possible biases. As for PROGRAPHER, since it is not open source, we reproduce it and will release its code. Our experiments were carried out on a server running Ubuntu 18.04 64-bit OS with Intel(R) Xeon(R) Silver 4214R CPU @ 2.40GHz, 256GB memory.
\subsection{Metrics}
We define window-level precision as: $\frac{W_{TP}}{W_{TP} + W_{FP}}$, where $W_{TP}$ and $W_{FP}$ are window-level true positives and false positives, respectively. We consider a provenance graph (time window) to be $W_{TP}$ if it contains attack steps and is reported as an alert. Otherwise, we consider it as $W_{FP}$. We define window-level recall as $\frac{W_{TP}}{W_{TP} + W_{FN}}$, where $W_{FN}$ is the number of window-level false negatives, which contain attack steps but are not identified as alerts. We define window-level accuracy as $Accuracy = \frac{W_{TP}+W_{TN}}{W_{TP}+W_{FP}+W_{TN}+W_{FN}}$ and window-level F1 as $F1 = 2* \frac{Precision*Recall}{Precision+Recall}$. We consider the definitions of window-level, graph-level, and node-level metrics to be similar, and to save space, we will not repeat the latter two.
\subsection{Datasets}
The purpose of \sn{} is to help long-running online detection systems combat the problem of false positives caused by concept drift that requires long-term datasets for evaluation. Given that there are multiple adversarial scenarios designed to mimic real-world Advanced Persistent Threats (APTs) on enterprise networks within the Transparent Computing (TC) program, we choose to employ datasets from DARPA's TC. Specifically, multiple red teams (e.g., CADETS, ClearScope, and THEIA) launched a series of attacks toward critical services such as web, email, and SSH servers, while engaging in benign activities such as browsing websites, checking emails, and SSH log-ins. The datasets from the third (E3) and fifth (E5) engagements of DARPA TC are publicly accessible~\cite{darpa3, darpa5}. 
\par
Table~\ref{tab:summaryExperDatasets} details the statistics from the different datasets, including duration time (specific dates and hours), the number of nodes and edges, the specific attack time and their proportions, as well as the number of attack-related nodes. The open-source report for DARPA TC E3~\cite{darpa3} and E5~\cite{darpa5} show an attack for E3-CLEARSCOPE on April 12, 2018 from 15:19 to 15:24, an attack for E5-CLEARSCOPE on May 17, 2019 from 15:42 to 16:00.  However, we couldn't find any relevant attack records (including file downloading and process creating) in the corresponding log files for this period. Therefore, we do not label it as attack time.

\begin{table*}[]
\centering
\caption{Summary of the experimental datasets. Columns 1 and 2 indicate the name and duration time of the corresponding dataset. Columns 3 and 4 indicate the number of nodes and edges. Columns 5 and 6 indicate the specific time of attack occurrence and the percentage of the attack time for the whole time.  Column 7 indicates the number of attack nodes involved in the attack window.}
\label{tab:summaryExperDatasets}
\resizebox{0.99\textwidth}{!}{%
\begin{tabular}{ccccccc}
\hline
\textbf{Dataset} & \textbf{Duration Time} & \textbf{\# N} & \textbf{\# E} & \textbf{Specific Attack Time}                       & \textbf{\% of Attack Time} & \textbf{\# of AN} \\ \hline
E3-CADETS        & 2018/4/2-13 (247h)     & 613,713       & 31,573,565    & 4.6 11:21-12:08 + 4.12 14:00-14:38 + 4.13 9:04-9:30 & 0.901\%                    & 49                \\
E3-ClearScope    & 2018/2-13 (168h)       & 369,021       & 22,254,213    & 4.11 13:55-14:47                                    & 0.892\%                    & 23                \\
E3-THEIA         & 2018/4/3-13 (142h)     & 1,258,362     & 44,366,117    & 4.10 13:41-14:55 + 4.12 12:44-13:26                 & 1.585\%                    & 97                \\
E5-CADETS        & 2019/5/8-17 (240h)     & 12,893,578    & 1,003,492,271 & 5:16 9:32-10:08 + 5.17 10:16-10:55                  & 0.625\%                    & 65                \\
E5-ClearScope    & 2019/5/8-17 (240h)     & 172,699       & 151,104,106   & 5.15 15:39-16:18                                    & 0.521\%                    & 81                \\
E5-THEIA         & 2019/5/8-15 (192h)     & 2,286,124     & 140,994,662   & 5.15 14:48-15:07                                    & 0.260\%                    & 36                \\ \hline
\textbf{Avg}     & /                      & 2,932,249     & 232,297,489   & /                                                   & 0.797\%                    & 58.5              \\ \hline
\end{tabular}%
}
\vspace{-0.5cm}
\end{table*}
\par
\textbf{Ground Truth Labeling.} 
In TC, attack activities occurred only in a subset of time windows within a specific day, as shown in Table~\ref{tab:summaryExperDatasets}. For instance, in our motivating example of Section~\ref{sec:2_motivation}, attackers executed \texttt{loaderDrakon} in the \texttt{Firefox} memory and escalated privileges for performing memory injection into the \texttt{sshd} to write file \texttt{sshdlog} between 14:58 and 15:07 on May 15, 2019. DARPA provides ground truth reports to aid us in quickly labeling. For Window-level, \sn{} partitions the streaming provenance logs into time windows and labels windows containing attack behaviors as \textbf{attack-related windows} while those lacking such behaviors as \textbf{benign windows}. For node-level, we manually identify all crucial nodes involved in attack behaviors as \textbf{attack-related nodes}, while labeling other nodes as \textbf{benign nodes}.

\subsection{RQ 1: The detection performance of \sn{}}
KAIROS~\cite{kairos} is a window-level detection system, THREATRACE~\cite{threatrace} is a node-level detection system, while UNICORN~\cite{han2020unicorn} and PROGRAPHER~\cite{prographer} are graph-level detection systems. To ensure fairness, we will compare \sn{} with three baselines at their respective levels to demonstrate the detection performance of \sn{}.
\subsubsection{Widnow-level}
\label{sec:6.4.1_windowsPerformance}
KAIROS detects anomalous windows based on the anomalousness and rareness of nodes. To ensure a fair comparison, we adopted the same time window size (15 minutes) and DARPA dataset for evaluating both KAIROS and \sn{} by calculating precision and recall at the window-level. 
\par
According to statistics shown in Table~\ref{tab:detection_performance_window_level}, \sn{} outperforms KAIROS comprehensively in both precision and recall at the window-level. For KAIROS, we replicated and evaluated its performance while \textbf{keeping its code and parameters unchanged}. However, its results were worse than those in the paper, which we attribute to three reasons after a detailed investigation. (1) We relabeled the datasets. For instance, in E3-CADETS, attacks occurred on three days, April 6, April 12, and April 13, involving a total of nine time windows. However, KAIROS only considered April 6 and ignored the attacks that occurred during the two days (April 12, and April 13). We relabeled the dataset and evaluated both \sn{} and KAIROS based on this. (2) KAIROS does not distinguish between nodes with the same name. For example, in E5-THEIA, there were a total of 6,647 unique IDs for the Firefox process, but KAIROS grouped them all under one single Firefox process. This causes KAIROS to convey node features incorrectly, which affects the judgment of anomalies. (3) KAIROS relies excessively on training data. KAIROS constructs an anomalous window queue by determining whether there are nodes with anomalousness (high reconstruction error) and rareness (high inverse document frequency, IDF) between different windows. The combination of the latter two factors results in KAIROS being unable to construct an anomalous window queue when attackers employ living-off-the-land attacks (involving no writing of files with special names to the disk). It is important to note that we only evaluated based on the new dataset labeling and did not modify the output of KAIROS.
\par
In terms of precision, \sn{} outperforms KAIROS by 30\%. As shown in Table~\ref{tab:detection_performance_window_level}, \sn{} generates an average of 6.7 false positives (FPs), which accounts for only 0.9\% of the average window count of 691 in the dataset. We believe this demonstrates \sn{}'s ability to efficiently mitigate alarm fatigue issues, allowing analysts to manage alerts more effectively. However, it is important to emphasize that \textbf{\sn{} only mitigates false positives caused by concept drift and does not eliminate all false alarms}. When the host's operating environment changes (such as switching work environments), \sn{} inevitably generates false alarms in the initial stages. As shown in Section~\ref{sec:4.2_overview}, we use a new anomaly detection paradigm to ensure that \sn{} ceases to generate alerts once the environment stabilizes after switching. But \sn{} will trigger alerts when users engage in rare behaviors (such as suddenly accessing unknown websites and downloading files). Given that such behaviors inherently share similarities with attack behaviors in terms of anomalousness and rareness, \sn{} does not consider handling such false positives to avoid false negatives. This is precisely why false alarms still exist in \sn{}. In Section~\ref{sec:6.5_componetsEffective}, we discuss in detail the importance of \sn{}'s components in mitigating false positives caused by concept drift.

\begin{table}[]
\centering
\caption{Experiment results for \sn{} and KAIROS at Window-level.}
\label{tab:detection_performance_window_level}
\resizebox{0.48\textwidth}{!}{%
\begin{tabular}{c|ccc|ccc}
\hline
\multirow{2}{*}{\textbf{Dataset}} & \multicolumn{3}{c|}{\textbf{KAIROS}} & \multicolumn{3}{c}{\textbf{\sn{}}}          \\
                                  & TP/FP/FN    & Precision   & Recall   & TP/FP/FN    & Precision     & Recall        \\ \hline
E3-CADETS                         & 7/9/2       & 0.44        & 0.88     & 11/4/0      & 0.73          & 1.00          \\
E3-ClearScope                     & 0/0/3       & 0.00        & 0.00     & 2/6/1       & 0.25          & 0.67          \\
E3-THEIA                          & 7/6/2       & 0.54        & 0.79     & 9/3/0       & 0.75          & 1.00          \\
E5-CADETS                         & 6/10/0      & 0.38        & 1.00     & 6/9/0       & 0.4           & 1.00          \\
E5-ClearScope                     & 3/18/0      & 0.14        & 1.00     & 3/13/0      & 0.18          & 1.00          \\
E5-THEIA                          & 1/1/1       & 0.50        & 0.50     & 2/5/0       & 0.29          & 1.00          \\ \hline
\textbf{Avg}                      & 4/7.3/1.3   & 0.33        & 0.66     & 5.5/6.7/0.2 & \textbf{0.43} & \textbf{0.95} \\ \hline
\end{tabular}%
}
\vspace{-0.3cm}
\end{table}

\par
In terms of Recall, \sn{} outperforms KAIROS by 41\%. Both \sn{} and KAIROS exhibit poor recall performance on the E3-CLEARSCOPE dataset (0 and 0.67, respectively). E3-ClearScope represents a vulnerability exploitation attack that occurred on the Android platform. On April 11th, the attacker gained control of Firefox by browsing the malicious website \texttt{www.mit.gov.jo}. Following that, the attacker downloaded a payload (\texttt{shared\_files}) and executed it with root privileges. Finally, several attempts were made to load modules and inject processes, but both failed. For unknown reasons, there is no associated process creation behavior (CLONE/FORK/EXECUTE) in ClearScope, which prevents the transmission of inter-process information, significantly impacting \sn{}'s recall performance. However, on the other datasets, \sn{} maintains a recall of 1. Therefore, we believe that \sn{} can detect attack-related windows when supported by a good auditing system (as described in Section~\ref{sec:4_threatmodel}).

\subsubsection{Graph-level} 
UNICORN transforms streaming data into fixed-size, updatable graph sketches that preserve Jaccard similarity~\cite{yang2017histosketch} to model a system's behavior. Each sketch represents a snapshot, depicting an overall summary of the system's state from the beginning of execution to the time of snapshot capture. PROGRAPHER~\cite{prographer}, on the other hand, converts snapshots into fixed-size vectors using graph neural networks graph2vec~\cite{narayanan2017graph2vec} and leverages recurrent neural networks TextRCNN~\cite{lai2015textrcnn} to learn benign system behavior changes.
\par
Considering both the above studies are graph-level detection systems, we employed the same evaluation protocol as theirs to guarantee fairness, i.e., determining whether the entire graph is benign or malicious. We classify a graph to be malicious if the time window of the \sn{} alert is a subset of the corresponding time window of the graph. Since UNICORN and PROGRAPHER were evaluated only on the E3 dataset (CADETS, ClearScope, THEIA), \sn{} was also compared on the same dataset to avoid the bias introduced by hyper-parameters tuning.
\par
\begin{figure}[h!t]
\centering
\includegraphics[width=\linewidth]{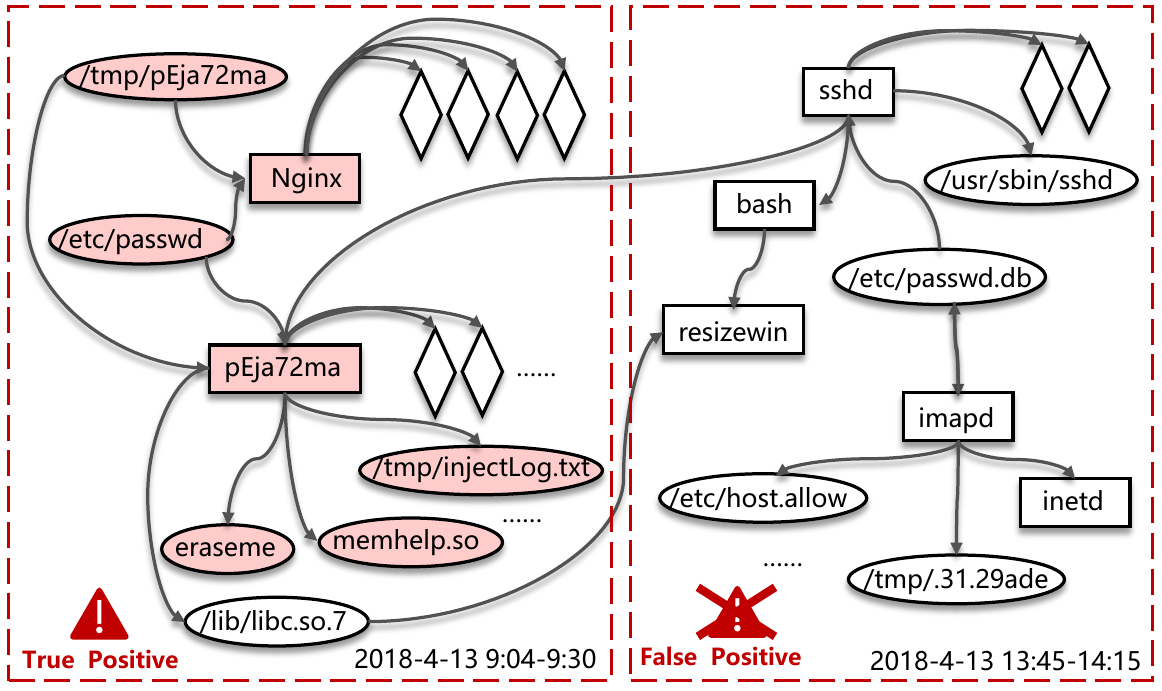}
\caption{Reconstructed Attack Scenario. The red dashed box indicates the window-level alerts generated by \sn{}, with the specific time shown in the bottom right corner. On the left are the true positive, and on the right are the false positive. Red nodes represent attack-related nodes, while white nodes represent non-attack-related nodes. 
}
\label{fig:reconstructed_attack_scenario}
\vspace{-0.3cm}
\end{figure}

The experiment results for \sn{} and the other two baselines at graph-level are reported in Table~\ref{tab:graph_performance}. \sn{} detects all attacks in three datasets (1.0 recall) and only identified one benign graph in E3-CADETS as malicious, indicating its effectiveness in capturing attack traces while avoiding alert fatigue issues. Further investigation revealed that for the E3-CADETS dataset, \sn{} incorrectly identified the window between 13:45 and 14:15 on 2018.4.13 as a malicious window. This was due to an \texttt{imapd} process being identified as a suspicious node during this period, and both the \texttt{imapd} process and \texttt{sshd} (parent process of malicious process \texttt{pEja72ma}) interacted with the same sensitive file, \texttt{/etc/spwd.db}, creating a suspicious path, as shown in Figure~\ref{fig:reconstructed_attack_scenario}. Additionally, this period is a subset of the period corresponding to a graph, leading to a graph-level false positive. However, in real-world scenarios, attackers may use an injected \texttt{sshd} to accomplish multiple tasks across periods. Therefore, this case requires security analysts' intervention to determine whether it is a false positive.
\par
For comparison, UNICORN and PROGRAPHER all perform worse than \sn{} in nearly every metric. We learned that there are two reasons why UNICORN's performance is worse than reported in the original paper: (1) UNICORN was trained using benign datasets under DARPA TC that were not publicly available, leading to bias in the model. (2) UNICORN did not enforce that the testing graphs occur after training and validation, which is a typical Data Snooping~\cite{arp2022datasnooping}. While PROGRAPHER was not open-sourced, we replicated it based on the description of the paper. Keeping the hyper-parameters and necessary components unchanged, we believe the main reason for PROGRAPHER's poorer performance compared to the original paper is that the encoder (Graph2Vec) cannot generate suitable embedding vectors. We primarily used Doc2Vec~\cite{le2014doc2vec} provided by Gensim~\cite{gensim} in combination with RSG mentioned in PROGRAPHER to generate embedding vectors for snapshots, but the performance was unsatisfactory. We attribute the result to our lack of expertise rather than the original authors, and we intend to open-source the reproduction code of PROGRAPHER.

\begin{table}[]
\centering
\caption{Experiment results for \sn{}, UNICORN and PROGRAPHER at Graph-level.}
\label{tab:graph_performance}
\resizebox{0.48\textwidth}{!}{%
\begin{tabular}{c|lllll}
\hline
\textbf{Dataset}               & \textbf{System} & \textbf{Precision} & \textbf{Recall} & \textbf{Accuray} & \textbf{F1}   \\ \hline
\multirow{3}{*}{E3-CADETS}     & UNICORN         & 0.31               & 1.00            & 0.44             & 0.47          \\
                               & PROGRAPHER      & 0.50               & 1.00            & 0.96             & 0.67          \\
                               & \sn{}           & \textbf{0.80}      & \textbf{1.00}   & \textbf{1.00}    & \textbf{0.88} \\ \hline
\multirow{3}{*}{E3-ClearScope} & UNICORN         & 1.0                & 0.75            & 0.93             & 0.89          \\
                               & PROGRAPHER      & 0.60               & 1.00            & 0.98             & 0.86          \\
                               & \sn{}           & \textbf{1.00}      & \textbf{1.00}   & \textbf{1.00}    & \textbf{1.00} \\ \hline
\multirow{3}{*}{E3-THEIA}      & UNICORN         & 0.67               & 0.67            & 0.8              & 0.67          \\
                               & PROGRAPHER      & 0.60               & 1.00            & 0.97             & 0.75          \\
                               & \sn{}           & \textbf{1.00}      & \textbf{1.00}   & \textbf{1.00}    & \textbf{1.00} \\ \hline
\end{tabular}%
}
\vspace{-0.3cm}
\end{table}

\subsubsection{Node-level} THREATRACE is a detection system that assesses the anomalousness of nodes using a multi-model framework. While \sn{} generates a reconstruction graph of the attack scenario corresponding to the malicious windows, allowing for evaluation at the node level for both. THREATRACE~\cite{threatrace} is an open-source system, enabling us to run it directly for comparison. To maintain fairness, we adopted THREATRACE's way of computing metrics and selected a subset of the DARPA TC dataset used by THREATRACE.
\par
As shown in Table~\ref{tab:node_performance}, \sn{} significantly outperforms THREATRACE in precision (improved by 29\%), while slightly lagging in recall (decreased by 4\%). Through our further investigation, we identified two main reasons for this discrepancy. Firstly, THREATRACE employs heuristic extensions to the node-level GroundTruth. Specifically, \textit{THREATRACE labels malicious nodes and their neighbors within two hops as GroundTruth, even if the neighbor nodes are not directly involved in the attack}. For instance, in E5-THEIA, the number of malicious nodes labeled by THREATRACE (162,714) is 4,519 times higher than that labeled by \sn{} (36). This GroundTruth leads to poorer recall performance for \sn{}. Secondly, THREATRACE discretely assesses node anomalies. This results in malicious nodes being identified across various windows without establishing their relationships. In contrast, \sn{} delineates malicious windows, merges malicious paths, identifies malicious nodes, and ultimately generates a reconstructed graph representing malicious behavior. This superior approach contributes to the higher precision of \sn{}, as all the malicious nodes it identifies (positives) are around real malicious nodes (true positives).
\par
In addition, THREATRACE cannot to reconstruct a complete attack scenario. Node-level information can indeed assist analysts in understanding attack behavior relative to graph-level data. However, when the number of false positive nodes is large, such as the 63,137 false positives in E5-THEIA for THREATRACE (under such lenient labeling conditions), security analysts are still inundated with alerts, making them difficult to take countmeasures timely.

\begin{table}[]
\centering
\caption{Experiment results for \sn{} and THREATRACE at Node-level.}
\label{tab:node_performance}
\begin{tabular}{c|llll}
\hline
\textbf{Dataset}           & \textbf{System} & \textbf{Precision} & \textbf{Recall} & \textbf{Accuracy} \\ \hline
\multirow{2}{*}{E3-CADETS} & THREATRACE      & 0.90               & 0.99            & 0.99              \\
                           & \sn{}           & \textbf{1.00}      & 0.92            & 0.98              \\ \hline
\multirow{2}{*}{E3-THEIA}  & THREATRACE      & 0.87               & 0.99            & 0.99              \\
                           & \sn{}           & \textbf{1.00}      & 0.94            & 0.99              \\ \hline
\multirow{2}{*}{E5-CADETS} & THREATRACE      & 0.63               & 0.86            & 0.97              \\
                           & \sn{}           & \textbf{1.00}      & 0.84            & \textbf{0.98}     \\ \hline
\multirow{2}{*}{E5-THEIA}  & THREATRACE      & 0.70               & 0.92            & 0.99              \\
                           & \sn{}           & \textbf{1.00}      & 0.90            & 0.98              \\ \hline
\end{tabular}
\end{table}

\subsection{RQ2: The Importance of \sn{}'s Components}
\label{sec:6.5_componetsEffective}
The goal of \sn{} is to mitigate the high false positives caused by concept drift in long-running anomaly detection systems. In other words, \textbf{\sn{} focuses on reducing false positives caused by concept drift rather than addressing all false positives.} Currently, \sn{} attempts to mitigate high false positives by introducing incremental learning as a new anomaly detection paradigm. However, there are some challenges, as described in Section~\ref{sec:1_intro}, for which we have designed specific components in \sn{} to address them. In this section, we focus on showing the performance and effectiveness of these components and answer questions step-by-step.
\par
\begin{figure}[h!t]
\centering
\includegraphics[width=0.37\textwidth]{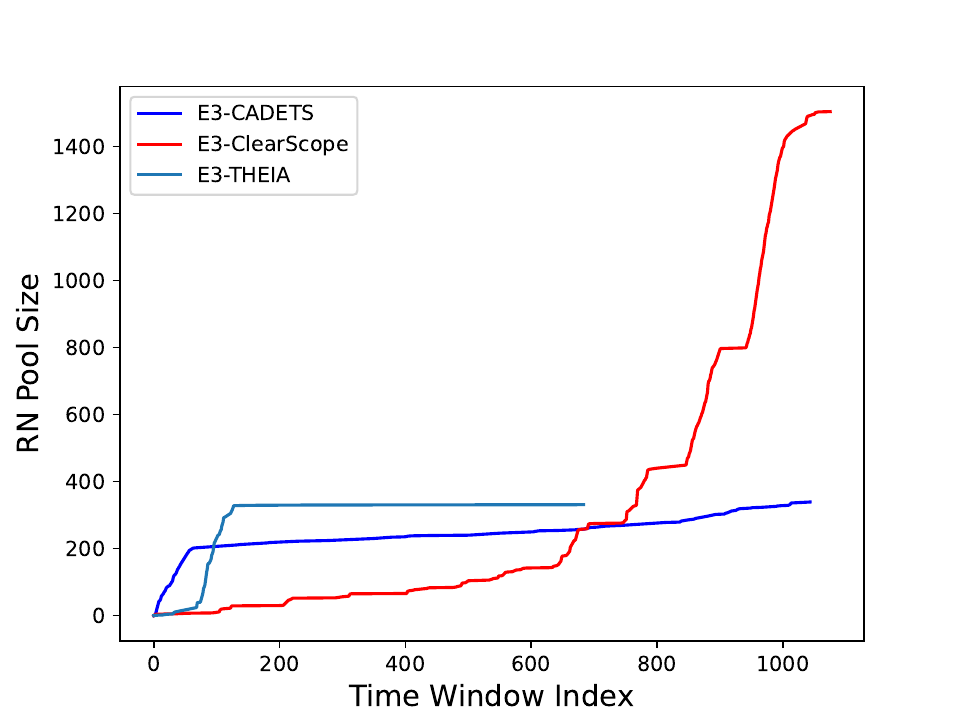}
\caption{Growth Trend of RN Pool size over time. 
}
\label{fig:rn_pool_size}
\vspace{-0.3cm}
\end{figure}

\textbf{How does the size of the RN Pool change over time?} Figure~\ref{fig:rn_pool_size} shows the variation trend of the RN Pool size in the corresponding dataset. It can be observed that the RN Pool corresponding to E3-ClearScope exhibits an abnormal growth rate, attributed to the absence of process creation-related events (FORK/CLONE). This results in \sn{} initializing all processes as benign without suspiciousness, and identifying numerous anomaly nodes as rehearsal nodes. In both E3-CADETS and E3-ClearScope, the RN Pool experiences rapid growth in the initial stages, with many nodes being identified as anomalies but benign, effectively aiding \sn{} in building a memory bank to swiftly reduce false alarms. Subsequently, the RN Pool experiences a slow increase as suspicious information propagates rapidly through event interactions, to avoid \sn{} memorizing malicious behaviors.

\par
\textbf{Can the RN Pool be contaminated?}
From Table~\ref{tab:rnpool_contamination}, we can observe that the constructed RN Pool has no intersection with the attack nodes across almost all datasets. This is attributed to \sn{}'s stringent suspicious state transfer rules in constructing the RN Pool, thereby enabling \sn{} to avoid learning malicious behaviors and introducing false negatives. Prior work~\cite{sleuth, xiong2020conan} suggests that suspicious state transfer captures the essence of intrusion attacks, and \sn{} utilizes this rule to inversely identify rehearsal nodes that can provide replay information of critical scenarios. Additionally, in the case of E3-ClearScope, due to the same reason (absence of events related to process creation), the suspicious state can't transfer effectively, resulting in all attack-related nodes being included in the RN Pool. We attribute this to issues with the auditing system.

\begin{table}[]
\centering
\caption{Contamination analysis of RN Pool.}
\label{tab:rnpool_contamination}
\resizebox{0.43\textwidth}{!}{%
\begin{tabular}{c|cc}
\hline
\textbf{Dataset}       & \textbf{RN Pool Size} & \textbf{\# of Attack Nodes (\%)} \\ \hline
E3-CADETS     & 339                   & 0 (0\%)                          \\
E3-ClearScope & 1504                  & 12 (0.8\%)                       \\
E3-THEIA      & 616                   & 0 (0\%)                          \\
E5-CADETS     & 832                   & 0 (0\%)                          \\
E5-ClearScope & 292                   & 0 (0\%)                          \\
E5-THEIA      & 331                   & 0 (0\%)                          \\ \hline
\end{tabular}%
}
\vspace{-0.5cm}
\end{table}
 
\par
\textbf{Why is KAIROS able to achieve similar precision with \sn{} at the window level?}
We have evaluated \sn{} and KAIROS at the window level in Section~\ref{sec:6.4.1_windowsPerformance}. While both exhibit similar precision, their mechanisms for achieving this are different. KAIROS directly determines whether there is an intersection of nodes between different windows that exhibit both anomaly (high reconstruction loss) and rareness (high IDF), thereby constructing an anomalous window queue. We believe that KAIROS can accurately identify these nodes for two reasons: (1) KAIROS aggregates all nodes with the same name into one node, and (2) KAIROS carefully selects the benchmark set for constructing IDF. Upon detailed analysis, we found that all nodes supporting the construction of the anomalous window queue in KAIROS are not present in the benchmark set for constructing IDF. We consider this method relies too heavily on the benchmark set and is overly sensitive. For E3-CADETS, KAIROS used the benchmark set (4.3, 4.4, and 4.5) to build the IDF and tested it on the test set (4.6 and 4.7). As shown in Table~\ref{tab:idf_fine_tuning}, when we fine-tuned the days of the IDF benchmark set, KAIROS produced a significant number of false positives, leading to a substantial decrease in precision. In contrast, \sn{} employs a new anomaly detection paradigm and does not suffer similar effects.

\vspace{0.5cm}
\begin{table}[]
\centering
\caption{Impact of fine-tuning the benchmark set for building IDF on KAIROS. In the second column, 4.3-5 refers to the selection of CADETS corresponding to data from April 3 to April 5, 2018, to construct IDF benchmark set.}
\label{tab:idf_fine_tuning}
\resizebox{0.48\textwidth}{!}{%
\begin{tabular}{c|c|ccccc}
\hline
\textbf{Dataset}                    & \textbf{\begin{tabular}[c]{@{}c@{}}Selected\\ Days\end{tabular}} & \textbf{TP/FP/FN} & \textbf{Precision} & \textbf{Recall} & \textbf{Accuracy} & \textbf{F1} \\ \hline
\multirow{3}{*}{\textbf{E3-CADETS}} & \textbf{\begin{tabular}[c]{@{}c@{}}IDF\\ 4.3-5\end{tabular}}     & 4/1/0             & 0.80               & 1.0             & 0.99              & 0.89        \\ \cline{2-7} 
                                    & \textbf{\begin{tabular}[c]{@{}c@{}}IDF\\ 4.2-4\end{tabular}}     & 4/95/80           & 0.04               & 1.0             & 0.47              & 0.08        \\ \cline{2-7} 
                                    & \textbf{\begin{tabular}[c]{@{}c@{}}IDF\\ 4.3-4\end{tabular}}     & 4/95/80           & 0.04               & 1.0             & 0.47              & 0.08        \\ \hline
\end{tabular}%
}
\end{table}

\par
\textbf{Can pseudo-edge connections, state transfer, and path-level filtering effectively reduce false positives?}
As mentioned in Section~\ref{sec:1_intro}, we believe that introducing incremental learning as a new paradigm for anomaly detection can mitigate the high false positives problem caused by concept drift. \sn{} combats catastrophic forgetting through pseudo-edge connections (Section~\ref{sec:5.2.1_pseudoEdgeConnection}), avoids discrimination paradox through state transfer (Section~\ref{sec:5.1.2_encoding}), and reduces typical false positives through path-level filtering (Section~\ref{sec:5.3.2_pathlevel_filtering}). As described in Table~\ref{tab:ablation_experiments}, we designed corresponding ablation experiments to demonstrate the effectiveness of these three components. It can be seen that incremental learning without any component is the most ineffective, with an average false positive count reaching 340 and an average precision of 0.018. Besides, the ablation of any component results in a significant deterioration in \sn{}'s performance, with an average decrease in precision rate of 81\% and an average increase of 81 false positives. This is because each component represents a critical solution proposed to address the significant challenges encountered in moving to a new paradigm of anomaly detection. These components do not operate independently, and only by complementing each other can achieve optimal results.

\begin{table*}[]
\centering
\caption{Ablation experiments on the effectiveness of \sn{}. 'Incremental Learning' only uses incremental learning as an anomaly detection paradigm. 'w/o Pseudo Edges' disables pseudo-edge connection. 'w/o State Transfer' disables state encoding and suspicious state judgment. 'w/o Path-level Filtering' disables path-leveling filtering.}
\label{tab:ablation_experiments}
\resizebox{0.95\textwidth}{!}{%
\begin{tabular}{c|cc|cc|cc|cc|cc}
\hline
\multirow{2}{*}{\textbf{Dataset}} & \multicolumn{2}{c|}{\textbf{\begin{tabular}[c]{@{}c@{}}Incremental \\ Learning\end{tabular}}} & \multicolumn{2}{c|}{\textbf{\begin{tabular}[c]{@{}c@{}}METANOIA\\ (w/o Pseudo Edges)\end{tabular}}} & \multicolumn{2}{c|}{\textbf{\begin{tabular}[c]{@{}c@{}}METANOIA\\ (w/o State Transfer)\end{tabular}}} & \multicolumn{2}{c|}{\textbf{\begin{tabular}[c]{@{}c@{}}METANOIA\\ (w/o Path Filtering)\end{tabular}}} & \multicolumn{2}{c}{\textbf{METANOIA}} \\ \cline{2-11} 
                                  & \textbf{Precision}                              & \textbf{Recall}                             & \textbf{Precision}                                 & \textbf{Recall}                                & \textbf{Precision}                                  & \textbf{Recall}                                 & \textbf{Precision}                                  & \textbf{Recall}                                 & \textbf{Precision}  & \textbf{Recall} \\ \hline
E3-CADETS                         & 0.025                                           & 1.00                                        & 0.09                                               & 1.00                                           & 0.16                                                & 1.00                                            & 0.11                                                & 1.00                                            & 0.73                & 1.00            \\
E3-ClearScope                     & 0.011                                           & 1.00                                        & 0.05                                               & 1.00                                           & 0.08                                                & 1.00                                            & 0.07                                                & 1.00                                            & 0.25                & 0.67            \\
E3-THEIA                          & 0.039                                           & 1.00                                        & 0.11                                               & 1.00                                           & 0.17                                                & 1.00                                            & 0.10                                                & 1.00                                            & 0.75                & 1.00            \\
E5-CADETS                         & 0.015                                           & 1.00                                        & 0.04                                               & 1.00                                           & 0.06                                                & 1.00                                            & 0.07                                                & 1.00                                            & 0.40                & 1.00            \\
E5-ClearScope                     & 0.010                                           & 1.00                                        & 0.03                                               & 1.00                                           & 0.05                                                & 1.00                                            & 0.05                                                & 1.00                                            & 0.18                & 1.00            \\
E5-THEIA                          & 0.008                                           & 1.00                                        & 0.06                                               & 1.00                                           & 0.10                                                & 1.00                                            & 0.12                                                & 1.00                                            & 0.29                & 1.00            \\ \hline
Avg                               & 0.018                                           & 1.00                                        & 0.06                                               & 1.00                                           & 0.10                                                & 1.00                                            & 0.09                                                & 1.00                                            & 0.43                & 0.95            \\ \hline
\end{tabular}%
}
\vspace{-0.3cm}
\end{table*}

\subsection{RQ3: The Effectiveness of Reconstructing Attack Scenarios}
Previous researches~\cite{kairos, alsaheel2021atlas, xu2022depcomm} indicate that reconstructing attack scenarios is an essential capability for existing detection systems. Unlike rule-based detection methods, anomaly-based detection systems focus more on this ability. The reason for this is that anomaly-based detection systems inherently have a certain amount of false positives, requiring involvement from security analysts. Reconstructing a complete and concise attack scenario can effectively assist security analysts in eliminating false positives and reducing time overhead.
\par
\sn{} has the capability to reconstruct real APT attack scenarios. Due to the operating system's use of multiple cloned subprocesses to collaboratively accomplish tasks, the attack scenario graph often contains numerous nodes with the same name. Therefore, to streamline the graph structure and facilitate better understanding by analysts, \sn{} merges nodes with the same name in the attack scenario graph to achieve optimal presentation. Table~\ref{tab:attackScenarioGraph} illustrates the sizes of the attack scenario graphs reconstructed by \sn{} from each dataset. We observe that the number of events identified as malicious windows is 18,041 times larger than the merged attack scenario graph. This demonstrates \sn{}'s ability to assist security analysts in rapidly comprehending attack scenarios, effectively filtering out false positives, and reducing the workload in \texttt{"}man in the loop\texttt{"}.

\begin{table}[]
\centering
\caption{Statistics of reconstructed attack scenario graphs. The second column is the total number of events in the window identified by \sn{} as malicious. The third column is the initial attack scenario graph. The fourth column is the attack scenario graph of merged nodes with the same name.}
\label{tab:attackScenarioGraph}
\begin{tabular}{c|c|cc|cc}
\hline
\multirow{2}{*}{\textbf{Dataset}} & \multirow{2}{*}{\textbf{\begin{tabular}[c]{@{}c@{}}\# of Edges in\\ Windows\end{tabular}}} & \multicolumn{2}{c|}{\textbf{Attack Graph}} & \multicolumn{2}{c}{\textbf{\begin{tabular}[c]{@{}c@{}}Attack Graph\\ (Merged)\end{tabular}}} \\ \cline{3-6} 
                                  &                                                                                            & \textbf{Nodes}       & \textbf{Edges}      & \textbf{Nodes}                                & \textbf{Edges}                               \\ \hline
E3-CADETS                         & 1,179,901                                                                                  & 426                  & 489                 & 101                                           & 119                                          \\
E3-ClearScope                     & 275,209                                                                                    & 217                  & 256                 & 41                                            & 47                                           \\
E3-THEIA                          & 1,485,706                                                                                  & 193                  & 241                 & 89                                            & 104                                          \\
E5-CADETS                         & 1,763,829                                                                                  & 504                  & 560                 & 106                                           & 122                                          \\
E5-ClearScope                     & 3,772,183                                                                                  & 698                  & 727                 & 143                                           & 159                                          \\
E5-THEIA                          & 1,128,869                                                                                  & 184                  & 213                 & 52                                            & 69                                           \\ \hline
Avg                               & 1,600,949                                                                                  & 370                  & 414                 & 87                                            & 103                                          \\ \hline
\end{tabular}
\vspace{-0.3cm}
\end{table}

\subsection{RQ4: The Influence of Hyper-parameters}
\begin{figure*}[t]
\centering
\subfloat[$\beta$]{\includegraphics[width=4.2cm]{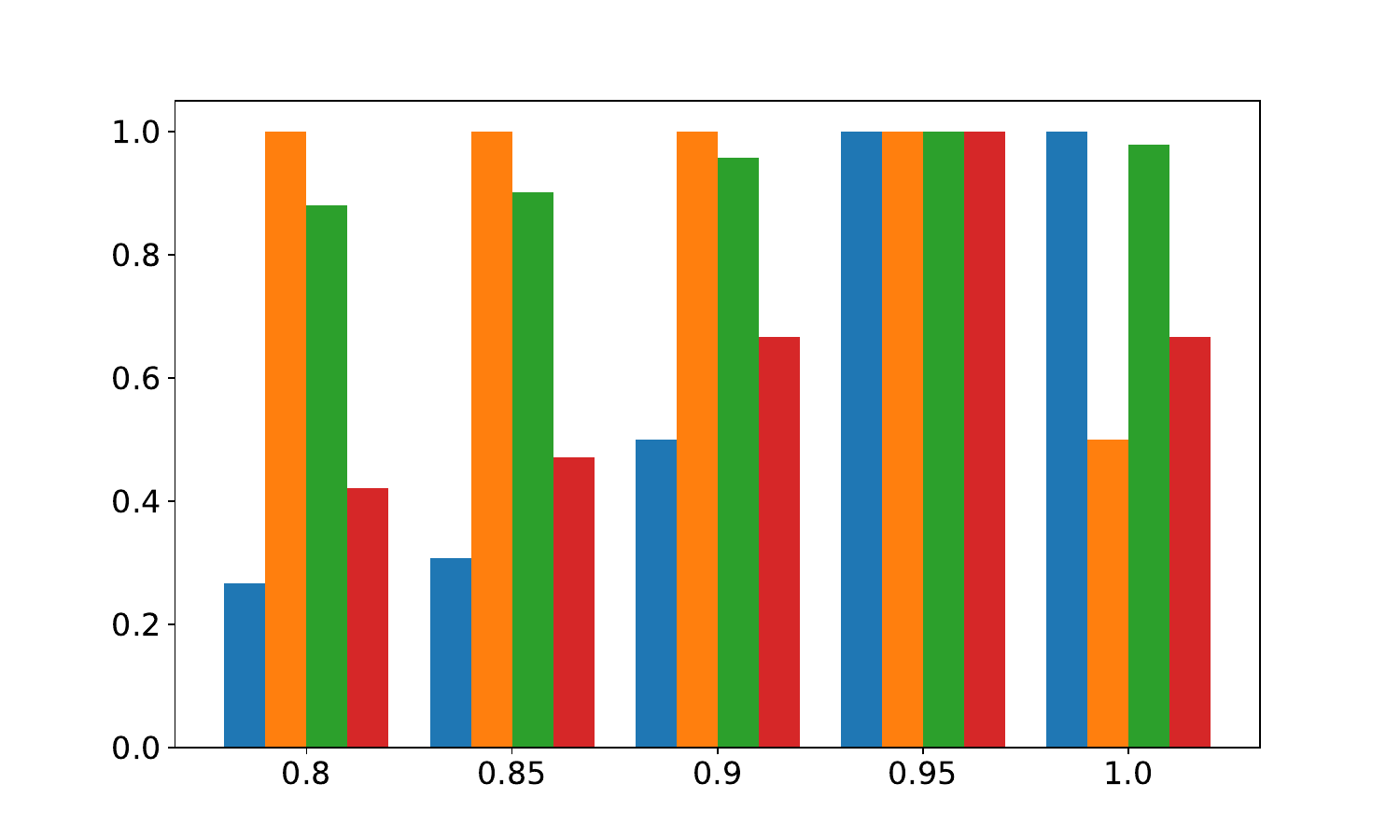}}
\subfloat[$\sigma$]{\includegraphics[width=4.2cm]{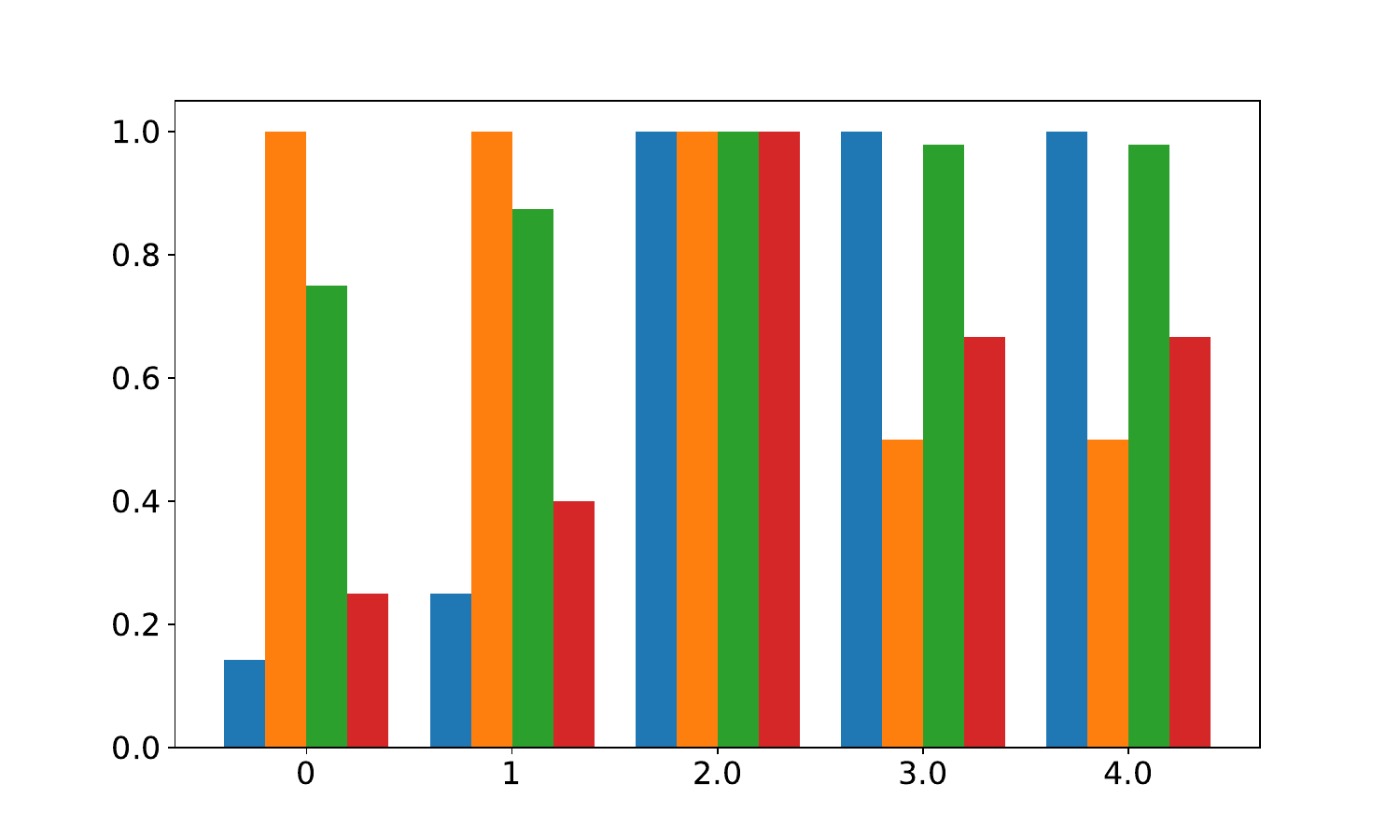}}
\subfloat[$\delta$]{\includegraphics[width=4.2cm]{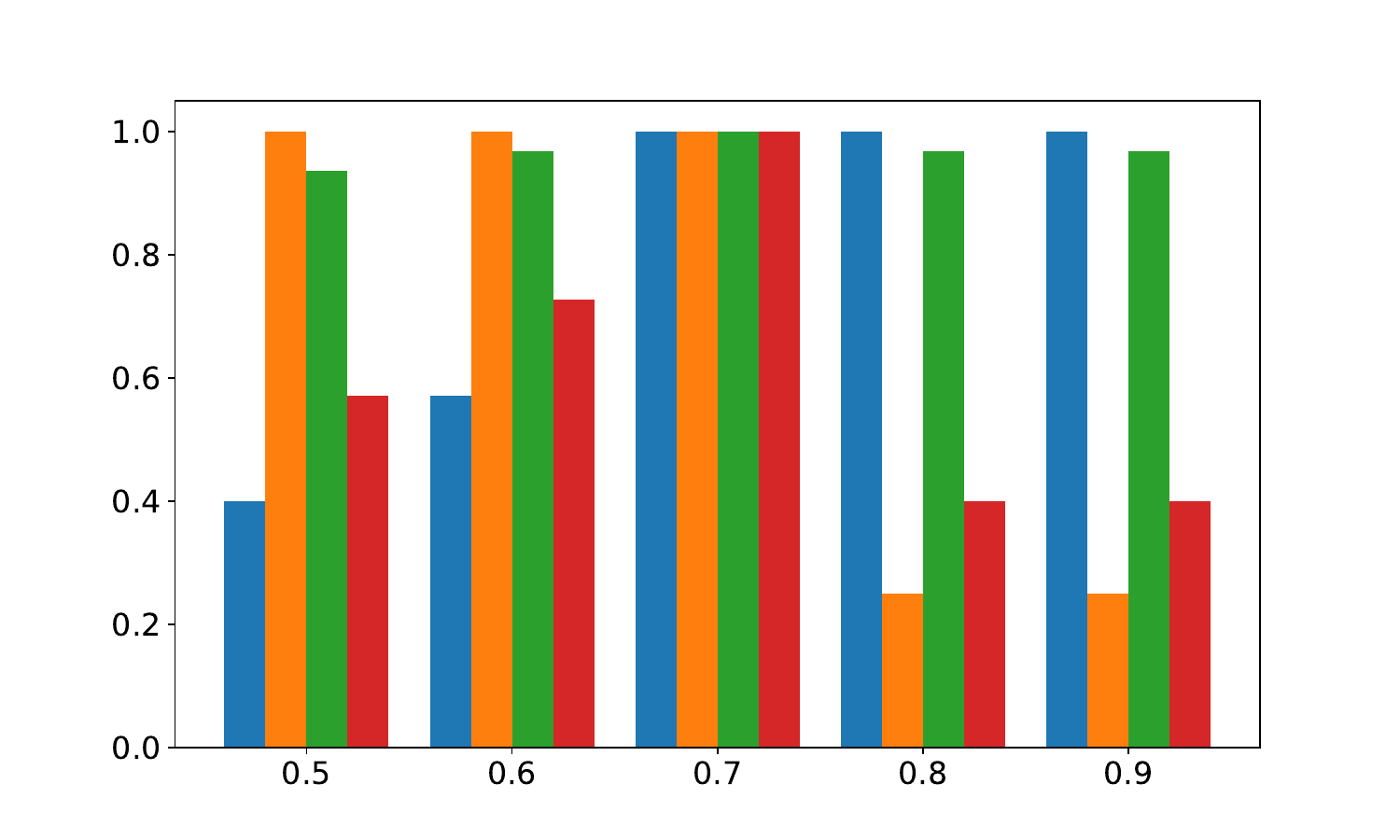}}
\subfloat[$\gamma$]{\includegraphics[width=4.2cm]{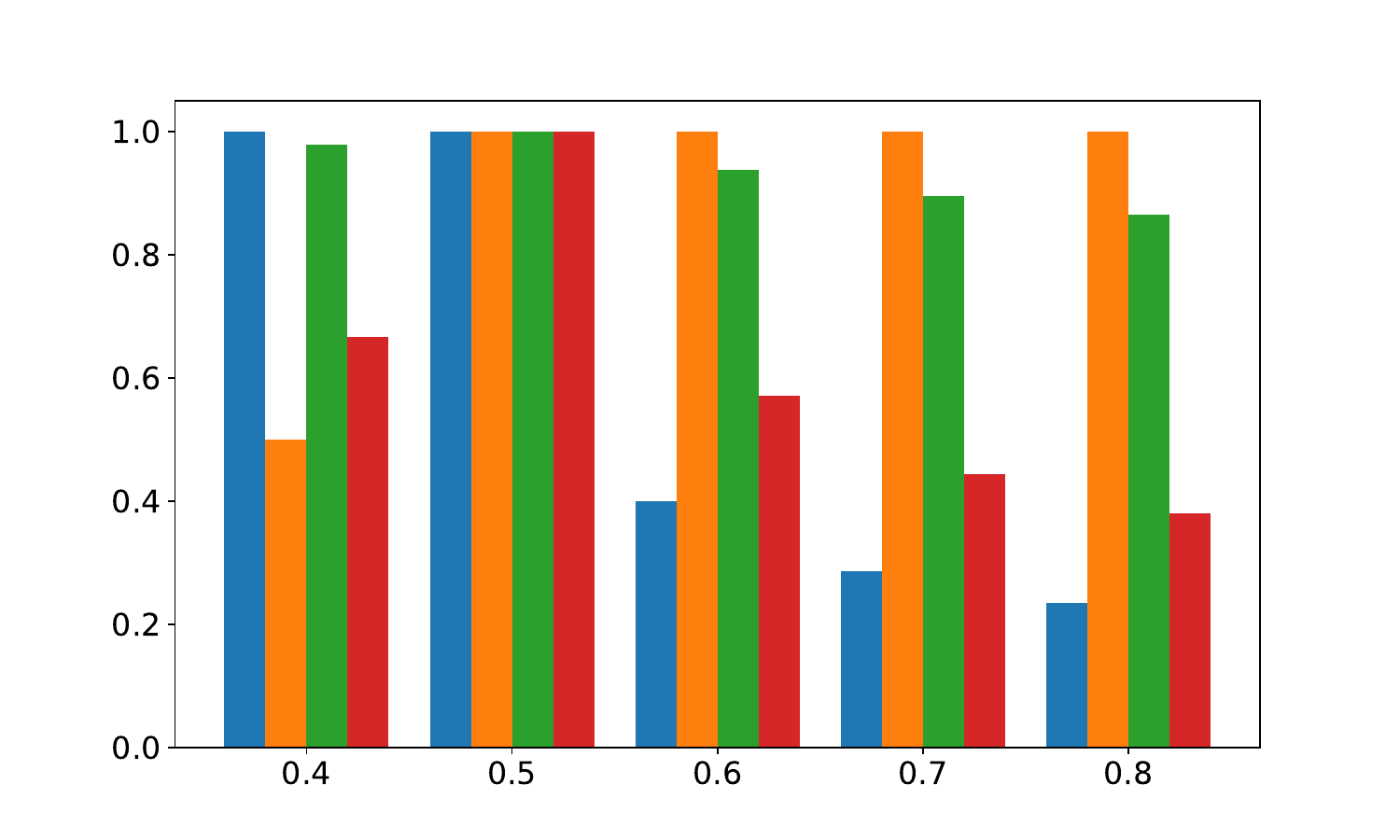}}
\caption{Impact of decay factor $\beta$, event anomaly threshold $\sigma$, path-level scoring threshold $\delta$ and node suspicion threshold $\gamma$. The four metrics corresponding to each parameter are, from left to right, Precision, Recall, Accuracy, and F1-Score.}
\label{fig:hyper-parameters}
\end{figure*}

We now analyze the impact of four key parameters on the effectiveness of \sn{}. Considering efficiency, we select only day 4.6 of the E3-CADETS dataset as the evaluation dataset. We independently vary the parameters to examine the influence of each parameter and the experimental results are shown in Figure~\ref{fig:hyper-parameters}.
\par
The decay factor $\beta$ refers to the decay rate of node suspicion. A smaller decay factor leads to a faster decrease in node suspicion, causing \sn{} to select more nodes. Besides, \sn{} determines if an event is anomalous through the event anomaly threshold $\sigma$, which is composed of the window's mean value of reconstruction loss plus variance. A higher abnormal threshold results in \sn{} retaining more anomaly events. Simultaneously, \sn{} determines if a path is anomalous through the path-level scoring threshold $\delta$. A higher $\delta$ leads to \sn{} retaining less anomalous paths. As illustrated in Figures 5(a), 5(b), and 5(c), with the increase of the decay factor, event anomaly threshold, and path-level scoring threshold, the precision of \sn{} increases while the recall decreases. At last, \sn{} determines if a node is suspicious through the node suspicion threshold $\gamma$. A higher $\gamma$ leads to \sn{} retaining more suspicious nodes. As shown in Figure 5(d), with the increase of the node suspicion threshold, the precision of \sn{} decreases while the recall increases.
\par
In the end, we choose the decay factor $\beta$ as 0.95, the event anomaly threshold $\sigma$ as mean reconstruction loss plus 2 standard deviations, the path-level scoring threshold $\delta$ as 0.7 and the node suspicion threshold $\gamma$ as 0.5.


\subsection{RQ5: The Overhead of \sn{}}
In this section, we evaluate the overhead of \sn{}, an online anomaly detection system that applies incremental learning. In this section, we set the time window to 15 minutes and measure the runtime overhead of each component, as shown in Table~\ref{tab:overhead}. When processing streaming logs, \sn{} requires an average of only 144.58 seconds (excluding state updates) to detect malicious windows and generate attack scenario graphs. This is significantly smaller than the duration of the window (15 minutes), demonstrating that \sn{} is a real-time anomaly detection system. Furthermore, considering extreme scenarios, we also evaluate the time overhead of \sn{} under the window with the most events. We selected the time window of 2019/5/15 23:41-23:56 under E5-CADETS, which contains 1,549,130 events, and \sn{} can complete the detection and investigation tasks within the corresponding time window, as shown in Table~\ref{tab:overhead}.

\begin{table}[]
\centering
\caption{Overhead of each component.}
\label{tab:overhead}
\resizebox{0.48\textwidth}{!}{%
\begin{tabular}{c|cc}
\hline
\textbf{Component}   & \textbf{Mean Duration} & \textbf{Max Duration} \\ \hline
Preprocess           & 16.71s                 & 93.37s                \\
Anomaly Detection    & 12.65s                 & 135.91s               \\
Malice Investigation & 42.93s                 & 107.12s               \\
State Update         & 62.34s                 & 509.03s               \\ \hline
Total                & 134.63s                & 745.43s               \\ \hline
\end{tabular}%
}
\vspace{-0.3cm}
\end{table}

\section{Discussion}
\label{sec:7_discussion}
\textbf{Evasion Attack.} Evasion attack refers to the actions taken by the attacker to mimic benign behavior to deceive detection systems during an attack. While evasion attacks pose a threat to all PIDSes, evading anomaly-based PIDSes is not straightforward. Therefore, to mimic benign behavior, attackers need to carefully orchestrate their attack actions, ensuring that malicious behavior exhibits benign features at the log level while maintaining their attack objectives. This requires attackers to have a deep understanding of both the target system's benign behavior and the mechanics of running PIDSes. Previous work~\cite{SIGL} has demonstrated that existing graph-based adversarial attacks cannot evade the detection of PIDSes because the provenance graphs have more structural features and time constraints. A recent robustness study~\cite{goyal2023sometimes} has indicated that evasion attacks can be achieved by automatically modifying system logs to evade existing PIDSes and have released evasion datasets based on DARPA E3-THEIA. However, KAIROS pointed out that this dataset only includes a small portion of the attack activities described in the DARPA ground truth, rather than the complete attack traces included in the original DARPA dataset. Further investigation confirmed this, and we did not proceed with further evaluation.
\par
\textbf{Poisoning Attack.} If an attacker can poison the training data to contain malicious activity for model learning, then future attacks will go undetected. To the best of our knowledge, SIGL~\cite{SIGL} is the only PIDS that evaluates its robustness against data poisoning in-depth, but only for the software installation scenario. ShadeWatcher~\cite{shadewatcher} inserts attack data from one day during training and demonstrates that it can still detect the corresponding attack. We performed a similar evaluation and also obtained good detection results. \sn{} adopts a new anomaly detection paradigm that requires replaying scenarios to combat catastrophic forgetting and is therefore more sensitive to poisoning attacks. \sn{} starts from the opposite side of the suspicious information flow to avoid the discrimination paradox and thus combat the poisoning attack.

\section{Related Work}
\label{sec:7_relatedwork}
In terms of detection methods, existing PIDSes can be categorized as heuristic-based and anomaly-based. Heuristic-based PIDSes~\cite{holmes, poirot, nodoze, rapsheet, sleuth} use empirical knowledge to construct matching rules to find known attacks in the provenance graph. However, they have difficulties in detecting unknown attacks and extending scale. Anomaly-based PIDSes detect intrusions by identifying deviations from normal behavior. Besides, offline PIDSes~\cite{shadewatcher, wang2020provdetector, SIGL}, due to their high operational overhead, fail to achieve efficient real-time intrusion detection. Online PIDSes~\cite{han2020unicorn, threatrace, prographer} relies on a fixed-size training set and cannot effectively address the problem of concept drift. UNICORN~\cite{han2020unicorn} combats concept drift by gradual forgetting but causes key information to decay with duration. KAIROS~\cite{kairos} relies on the intervention of analysts to re-train the model to combat concept drift, resulting in breaking attack-agnostic and high training costs. Although KAIROS acknowledges this issue in the paper, it does not provide a targeted solution, leaving it for future work.
\par
Moreover, previous PIDSes~\cite{pasquier2017practical, FRAPpuccino, streamspot, han2020unicorn} required manual investigation of large-scale provenance graphs, rendering them impractical for real-world use. Recently, THREATRACE~\cite{threatrace} and SHADEWATCHER~\cite{shadewatcher} began identifying anomalies at the node level, but they cannot reconstruct complete attack scenarios like \sn{}. SIGL~\cite{SIGL} attempts to generate graphs for software installation, but it struggles to scale to full-system intrusion detection. PROGRAPHER~\cite{prographer}, on the other hand, sorts the root subgraphs (RSGs) in snapshots by anomaly severity to help narrow down the scope of the investigation but loses the connection between multiple snapshots. Meanwhile, KAIROS~\cite{kairos} attempts to reconstruct attack scenarios in anomaly queues using community discovery algorithms but can only generate multiple star-shaped small attack scenarios (according to its open-source results).
\par
In conclusion, \sn{} is the first intrusion detection system for concept drift mitigation adapting a new anomaly detection paradigm with the ability to reconstruct concise and effective attack scenario graphs.

\section{Conclusion}
\label{sec:8_conclusion}
\sn{} is the first lifelong anomaly detection and investigation system that utilizes incremental learning as a new paradigm to address the high false positive problem caused by concept drift. 
Our evaluations demonstrate that \sn{} can monitor systems over long periods and achieve optimal performance.

\balance

\bibliographystyle{IEEEtran}
\bibliography{arxiv}

@misc{pyg,
  title = {PyTorch Geometric},
  year = 2024,
  note={\url{https://pytorch-geometric.readthedocs.io/en/latest/index.html.}},
}

@misc{gensim,
  title = {GENSIM:Topic Modelling for Humans},
  year = 2024,
  note={\url{https://radimrehurek.com/gensim/.}},
}

@inproceedings{goyal2023sometimes,
  title={Sometimes, you aren’t what you do: Mimicry attacks against provenance graph host intrusion detection systems},
  author={Goyal, Akul and Han, Xueyuan and Wang, Gang and Bates, Adam},
  booktitle={30th Network and Distributed System Security Symposium},
  year={2023}
}

@inproceedings{alsaheel2021atlas,
  title={$\{$ATLAS$\}$: A sequence-based learning approach for attack investigation},
  author={Alsaheel, Abdulellah and Nan, Yuhong and Ma, Shiqing and Yu, Le and Walkup, Gregory and Celik, Z Berkay and Zhang, Xiangyu and Xu, Dongyan},
  booktitle={30th USENIX security symposium (USENIX security 21)},
  pages={3005--3022},
  year={2021}
}

@inproceedings{le2014doc2vec,
  title={Distributed representations of sentences and documents},
  author={Le, Quoc and Mikolov, Tomas},
  booktitle={International conference on machine learning},
  pages={1188--1196},
  year={2014},
  organization={PMLR}
}

@misc{networkx,
  title = {NetworkX:Network Analysis in Python},
  year = 2024,
  note={\url{https://networkx.org/.}},
}

@misc{graphviz,
  title = {Graphviz},
  year = 2024,
  note={\url{https://graphviz.org/.}},
}

@article{kairos ,
  title={Kairos: Practical Intrusion Detection and Investigation using Whole-system Provenance},
  author={Zijun Cheng and Qiujian Lv and Jinyuan Liang and Yan Wang and Degang Sun and Thomas Pasquier and Xueyuan Han},
  journal={ArXiv},
  year={2023},
  volume={abs/2308.05034},
  url={https://api.semanticscholar.org/CorpusID:260735638}
}

@article{hwang1992steiner,
  title={Steiner tree problems},
  author={Hwang, Frank K and Richards, Dana S},
  journal={Networks},
  volume={22},
  number={1},
  pages={55--89},
  year={1992},
  publisher={Wiley Online Library}
}

@article{imase1991dynamicsteiner,
  title={Dynamic Steiner tree problem},
  author={Imase, Makoto and Waxman, Bernard M},
  journal={SIAM Journal on Discrete Mathematics},
  volume={4},
  number={3},
  pages={369--384},
  year={1991},
  publisher={SIAM}
}

@inproceedings{pasquier2017practical,
  title={Practical whole-system provenance capture},
  author={Pasquier, Thomas and Han, Xueyuan and Goldstein, Mark and Moyer, Thomas and Eyers, David and Seltzer, Margo and Bacon, Jean},
  booktitle={Proceedings of the 2017 Symposium on Cloud Computing},
  pages={405--418},
  year={2017}
}

@inproceedings{arp2022datasnooping,
  title={Dos and don'ts of machine learning in computer security},
  author={Arp, Daniel and Quiring, Erwin and Pendlebury, Feargus and Warnecke, Alexander and Pierazzi, Fabio and Wressnegger, Christian and Cavallaro, Lorenzo and Rieck, Konrad},
  booktitle={31st USENIX Security Symposium (USENIX Security 22)},
  pages={3971--3988},
  year={2022}
}

@inproceedings{yang2017histosketch,
  title={Histosketch: Fast similarity-preserving sketching of streaming histograms with concept drift},
  author={Yang, Dingqi and Li, Bin and Rettig, Laura and Cudr{\'e}-Mauroux, Philippe},
  booktitle={2017 IEEE International Conference on Data Mining (ICDM)},
  pages={545--554},
  year={2017},
  organization={IEEE}
}

@inproceedings{lai2015textrcnn,
  title={Recurrent convolutional neural networks for text classification},
  author={Lai, Siwei and Xu, Liheng and Liu, Kang and Zhao, Jun},
  booktitle={Proceedings of the AAAI conference on artificial intelligence},
  volume={29},
  number={1},
  year={2015}
}

@article{narayanan2017graph2vec,
  title={graph2vec: Learning distributed representations of graphs},
  author={Narayanan, Annamalai and Chandramohan, Mahinthan and Venkatesan, Rajasekar and Chen, Lihui and Liu, Yang and Jaiswal, Shantanu},
  journal={arXiv preprint arXiv:1707.05005},
  year={2017}
}

@article{he2011incremental,
  title={Incremental learning from stream data},
  author={He, Haibo and Chen, Sheng and Li, Kang and Xu, Xin},
  journal={IEEE Transactions on Neural Networks},
  volume={22},
  number={12},
  pages={1901--1914},
  year={2011},
  publisher={IEEE}
}

@article{van2022threeincremental,
  title={Three types of incremental learning},
  author={van de Ven, Gido M and Tuytelaars, Tinne and Tolias, Andreas S},
  journal={Nature Machine Intelligence},
  volume={4},
  number={12},
  pages={1185--1197},
  year={2022},
  publisher={Nature Publishing Group UK London}
}

@article{han2020unicorn,
  title={Unicorn: Runtime provenance-based detector for advanced persistent threats},
  author={Han, Xueyuan and Pasquier, Thomas and Bates, Adam and Mickens, James and Seltzer, Margo},
  journal={arXiv preprint arXiv:2001.01525},
  year={2020}
}

@inproceedings{holmes,
  title={HOLMES: real-time APT detection through correlation of suspicious information flows},
  author={Milajerdi, Sadegh M. and Gjomemo, Rigel and Eshete, Birhanu and Sekar, R. and Venkatakrishnan, V.N.},
  booktitle={2019 IEEE Symposium on Security and Privacy (SP)},
  pages={1137--1152},
  year={2019},
  organization={IEEE}
}

@inproceedings{poirot,
  title={Poirot: Aligning attack behavior with kernel audit records for cyber threat hunting},
  author={S. M. Milajerdi and B. Eshete and R. Gjomemo and V. Venkatakrishnan},
  booktitle = {Proceedings of the 2019 ACM SIGSAC Conference on Computer and Communications Security},
  pages={1795-–1812},
  year={2019}
}

@article{nodoze,
  title={NoDoze: Combatting Threat Alert Fatigue with Automated Provenance Triage},
  author={Wajih Ul Hassan and Shengjian Guo and Ding Li and Zhengzhang Chen and Kangkook Jee and Zhichun Li and Adam Bates},
  journal={Proceedings 2019 Network and Distributed System Security Symposium},
  year={2019},
  url={https://api.semanticscholar.org/CorpusID:56436113}
}

@INPROCEEDINGS{rapsheet,
  author={Hassan, Wajih Ul and Bates, Adam and Marino, Daniel},
  booktitle={2020 IEEE Symposium on Security and Privacy (SP)}, 
  title={Tactical Provenance Analysis for Endpoint Detection and Response Systems}, 
  year={2020},
  pages={1172-1189}
}

@inproceedings{sleuth,
  author = {Hossain, Md Nahid and Milajerdi, Sadegh M. and Wang, Junao and Eshete, Birhanu and Gjomemo, Rigel and Sekar, R. and Stoller, Scott D. and Venkatakrishnan, V. N.},
  title = {SLEUTH: Real-Time Attack Scenario Reconstruction from COTS Audit Data},
  year = {2017},
  booktitle = {Proceedings of the 26th USENIX Conference on Security Symposium},
  pages = {487–504}
}

@INPROCEEDINGS{shadewatcher,
  author={Zengy, Jun and Wang, Xiang and Liu, Jiahao and Chen, Yinfang and Liang, Zhenkai and Chua, Tat-Seng and Chua, Zheng Leong},
  booktitle={2022 IEEE Symposium on Security and Privacy (SP)}, 
  title={SHADEWATCHER: Recommendation-guided Cyber Threat Analysis using System Audit Records}, 
  year={2022},
  pages={489-506}
}

@ARTICLE{threatrace,
  author={Wang, Su and Wang, Zhiliang and Zhou, Tao and Sun, Hongbin and Yin, Xia and Han, Dongqi and Zhang, Han and Shi, Xingang and Yang, Jiahai},
  journal={IEEE Transactions on Information Forensics and Security}, 
  title={THREATRACE: Detecting and Tracing Host-Based Threats in Node Level Through Provenance Graph Learning}, 
  year={2022},
  pages={3972-3987}
}

@article{gama2014conceptdriftsurvey,
  title={A survey on concept drift adaptation},
  author={Gama, Jo{\~a}o and {\v{Z}}liobait{\.e}, Indr{\.e} and Bifet, Albert and Pechenizkiy, Mykola and Bouchachia, Abdelhamid},
  journal={ACM computing surveys (CSUR)},
  volume={46},
  number={4},
  pages={1--37},
  year={2014},
  publisher={ACM New York, NY, USA}
}

@inproceedings{pasquier2017camflow,
  title={Practical whole-system provenance capture},
  author={Pasquier, Thomas and Han, Xueyuan and Goldstein, Mark and Moyer, Thomas and Eyers, David and Seltzer, Margo and Bacon, Jean},
  booktitle={Proceedings of the 2017 Symposium on Cloud Computing},
  pages={405--418},
  year={2017}
}

@inproceedings{dong2023we,
  title={Are we there yet? An Industrial Viewpoint on Provenance-based Endpoint Detection and Response Tools},
  author={Dong, Feng and Li, Shaofei and Jiang, Peng and Li, Ding and Wang, Haoyu and Huang, Liangyi and Xiao, Xusheng and Chen, Jiedong and Luo, Xiapu and Guo, Yao and others},
  booktitle={Proceedings of the 2023 ACM SIGSAC Conference on Computer and Communications Security},
  pages={2396--2410},
  year={2023}
}

@inproceedings{ma2017mpi,
  title={$\{$MPI$\}$: Multiple perspective attack investigation with semantic aware execution partitioning},
  author={Ma, Shiqing and Zhai, Juan and Wang, Fei and Lee, Kyu Hyung and Zhang, Xiangyu and Xu, Dongyan},
  booktitle={26th USENIX Security Symposium (USENIX Security 17)},
  pages={1111--1128},
  year={2017}
}

@inproceedings{paccagnella2020logging,
  title={Logging to the danger zone: Race condition attacks and defenses on system audit frameworks},
  author={Paccagnella, Riccardo and Liao, Kevin and Tian, Dave and Bates, Adam},
  booktitle={Proceedings of the 2020 ACM SIGSAC Conference on Computer and Communications Security},
  pages={1551--1574},
  year={2020}
}

@inproceedings{yagemann2021validating,
  title={Validating the integrity of audit logs against execution repartitioning attacks},
  author={Yagemann, Carter and Noureddine, Mohammad A and Hassan, Wajih Ul and Chung, Simon and Bates, Adam and Lee, Wenke},
  booktitle={Proceedings of the 2021 ACM SIGSAC Conference on Computer and Communications Security},
  pages={3337--3351},
  year={2021}
}

@inproceedings{paccagnella2020custos,
  title={Custos: Practical tamper-evident auditing of operating systems using trusted execution},
  author={Paccagnella, Riccardo and Datta, Pubali and Hassan, Wajih Ul and Bates, Adam and Fletcher, Christopher and Miller, Andrew and Tian, Dave},
  booktitle={Network and distributed system security symposium},
  year={2020}
}

@inproceedings{karande2017sgx,
  title={SGX-Log: Securing system logs with SGX},
  author={Karande, Vishal and Bauman, Erick and Lin, Zhiqiang and Khan, Latifur},
  booktitle={Proceedings of the 2017 ACM on Asia Conference on Computer and Communications Security},
  pages={19--30},
  year={2017}
}

@inproceedings{lee2013loggc,
  title={Loggc: garbage collecting audit log},
  author={Lee, Kyu Hyung and Zhang, Xiangyu and Xu, Dongyan},
  booktitle={Proceedings of the 2013 ACM SIGSAC conference on Computer \& communications security},
  pages={1005--1016},
  year={2013}
}

@article{faber2023lifelongforanomalydetection,
  title={Lifelong Learning for Anomaly Detection: New Challenges, Perspectives, and Insights},
  author={Faber, Kamil and Corizzo, Roberto and Sniezynski, Bartlomiej and Japkowicz, Nathalie},
  journal={arXiv preprint arXiv:2303.07557},
  year={2023}
}

@misc{darpa5,
  year = 2020,
  title = {Transparent Computing Engagement 5 Data Release},
  note={\url{ https://github.com/darpa-i2o/Transparent-Computing}},
  author = {J. Torrey},
}

@article{jia2023magic,
  title={MAGIC: Detecting Advanced Persistent Threats via Masked Graph Representation Learning},
  author={Jia, Zian and Xiong, Yun and Nan, Yuhong and Zhang, Yao and Zhao, Jinjing and Wen, Mi},
  journal={arXiv preprint arXiv:2310.09831},
  year={2023}
}

@article{chen2022aptkgl,
  title={APT-KGL: An Intelligent APT Detection System Based on Threat Knowledge and Heterogeneous Provenance Graph Learning},
  author={Chen, Tieming and Dong, Chengyu and Lv, Mingqi and Song, Qijie and Liu, Haiwen and Zhu, Tiantian and Xu, Kang and Chen, Ling and Ji, Shouling and Fan, Yuan},
  journal={IEEE Transactions on Dependable and Secure Computing},
  year={2022},
  publisher={IEEE}
}

@article{febrinanto2023lifelongsurvey,
  title={Graph lifelong learning: A survey},
  author={Febrinanto, Falih Gozi and Xia, Feng and Moore, Kristen and Thapa, Chandra and Aggarwal, Charu},
  journal={IEEE Computational Intelligence Magazine},
  volume={18},
  number={1},
  pages={32--51},
  year={2023},
  publisher={IEEE}
}

@article{zipperle2022provenance,
  title={Provenance-based intrusion detection systems: A survey},
  author={Zipperle, Michael and Gottwalt, Florian and Chang, Elizabeth and Dillon, Tharam},
  journal={ACM Computing Surveys},
  volume={55},
  number={7},
  pages={1--36},
  year={2022},
  publisher={ACM New York, NY}
}

@article{lu2018conceptdriftlearning,
  title={Learning under concept drift: A review},
  author={Lu, Jie and Liu, Anjin and Dong, Fan and Gu, Feng and Gama, Joao and Zhang, Guangquan},
  journal={IEEE transactions on knowledge and data engineering},
  volume={31},
  number={12},
  pages={2346--2363},
  year={2018},
  publisher={IEEE}
}

@article{tsymbal2004conceptdrift,
  title={The problem of concept drift: definitions and related work},
  author={Tsymbal, Alexey},
  journal={Computer Science Department, Trinity College Dublin},
  volume={106},
  number={2},
  pages={58},
  year={2004},
  publisher={Dublin, Ireland}
}

@inproceedings {SIGL,
  author = {Xueyuan Han and Xiao Yu and Thomas Pasquier and Ding Li and Junghwan Rhee and James Mickens and Margo Seltzer and Haifeng Chen},
  title = {{SIGL}: Securing Software Installations Through Deep Graph Learning},
  booktitle = {30th USENIX Security Symposium (USENIX Security 21)},
  year = {2021},
  pages = {2345--2362},
  url = {https://www.usenix.org/conference/usenixsecurity21/presentation/han-xueyuan}
}

@inproceedings{wang2020provdetector,
  title={You Are What You Do: Hunting Stealthy Malware via Data Provenance Analysis.},
  author={Wang, Qi and Hassan, Wajih Ul and Li, Ding and Jee, Kangkook and Yu, Xiao and Zou, Kexuan and Rhee, Junghwan and Chen, Zhengzhang and Cheng, Wei and Gunter, Carl A and others},
  booktitle={NDSS},
  year={2020}
}

@article{xie2018pagoda,
  title={Pagoda: A hybrid approach to enable efficient real-time provenance based intrusion detection in big data environments},
  author={Xie, Yulai and Feng, Dan and Hu, Yuchong and Li, Yan and Sample, Staunton and Long, Darrell},
  journal={IEEE Transactions on Dependable and Secure Computing},
  volume={17},
  number={6},
  pages={1283--1296},
  year={2018},
  publisher={IEEE}
}

@inproceedings{FRAPpuccino,
  author = {Han, Xueyuan and Pasquier, Thomas and Ranjan, Tanvi and Goldstein, Mark and Seltzer, Margo},
  title = {FRAPpuccino: Fault-Detection through Runtime Analysis of Provenance},
  year = {2017},
  booktitle = {Proceedings of the 9th USENIX Conference on Hot Topics in Cloud Computing},
  pages = {18}
}

@inproceedings{morse,
  author={Hossain, Md Nahid and Sheikhi, Sanaz and Sekar, R.},
  booktitle={2020 IEEE Symposium on Security and Privacy (SP)}, 
  title={Combating Dependence Explosion in Forensic Analysis Using Alternative Tag Propagation Semantics}, 
  year={2020},
  volume={},
  number={},
  pages={1139-1155},
  doi={10.1109/SP40000.2020.00064}}

@inproceedings{streamspot,
  author = {Manzoor, Emaad and Milajerdi, Sadegh M. and Akoglu, Leman},
  title = {Fast Memory-Efficient Anomaly Detection in Streaming Heterogeneous Graphs},
  year = {2016},
  url = {https://doi.org/10.1145/2939672.2939783},
  booktitle = {Proceedings of the 22nd ACM SIGKDD International Conference on Knowledge Discovery and Data Mining},
  pages = {1035–1044}
}

@inproceedings{prographer,
author = {Yang, Fan and Xu, Jiacen and Xiong, Chunlin and Li, Zhou and Zhang, Kehuan},
title = {PROGRAPHER: An Anomaly Detection System Based on Provenance Graph Embedding},
year = {2023},
booktitle = {Proceedings of the 32nd USENIX Conference on Security Symposium}
}

@inproceedings{xu2016cpr,
  title={High fidelity data reduction for big data security dependency analyses},
  author={Xu, Zhang and Wu, Zhenyu and Li, Zhichun and Jee, Kangkook and Rhee, Junghwan and Xiao, Xusheng and Xu, Fengyuan and Wang, Haining and Jiang, Guofei},
  booktitle={Proceedings of the 2016 ACM SIGSAC conference on computer and communications security},
  pages={504--516},
  year={2016}
}

@inproceedings{king2003backtracking,
  title={Backtracking intrusions},
  author={King, Samuel T and Chen, Peter M},
  booktitle={Proceedings of the nineteenth ACM symposium on Operating systems principles},
  pages={223--236},
  year={2003}
}

@misc{DARPA-TC,
  year = 2020, 
  title = {DARPA Transparent Computing Engagement},
  note={\url{https://www.darpa.mil/program/transparent-computing.}},
}

@article{shi2020masked,
  title={Masked label prediction: Unified message passing model for semi-supervised classification},
  author={Shi, Yunsheng and Huang, Zhengjie and Feng, Shikun and Zhong, Hui and Wang, Wenjin and Sun, Yu},
  journal={arXiv preprint arXiv:2009.03509},
  year={2020}
}

@article{cho2014gru,
  title={Learning phrase representations using RNN encoder-decoder for statistical machine translation},
  author={Cho, Kyunghyun and Van Merri{\"e}nboer, Bart and Gulcehre, Caglar and Bahdanau, Dzmitry and Bougares, Fethi and Schwenk, Holger and Bengio, Yoshua},
  journal={arXiv preprint arXiv:1406.1078},
  year={2014}
}

@article{rossi2020temporal,
  title={Temporal graph networks for deep learning on dynamic graphs},
  author={Rossi, Emanuele and Chamberlain, Ben and Frasca, Fabrizio and Eynard, Davide and Monti, Federico and Bronstein, Michael},
  journal={arXiv preprint arXiv:2006.10637},
  year={2020}
}

@inproceedings{he2016resnet,
  title={Deep residual learning for image recognition},
  author={He, Kaiming and Zhang, Xiangyu and Ren, Shaoqing and Sun, Jian},
  booktitle={Proceedings of the IEEE conference on computer vision and pattern recognition},
  pages={770--778},
  year={2016}
}

@article{li2023nodlink,
  title={NODLINK: An Online System for Fine-Grained APT Attack Detection and Investigation},
  author={Li, Shaofei and Dong, Feng and Xiao, Xusheng and Wang, Haoyu and Shao, Fei and Chen, Jiedong and Guo, Yao and Chen, Xiangqun and Li, Ding},
  journal={arXiv preprint arXiv:2311.02331},
  year={2023}
}

@inproceedings{zhang2020dynamic,
  title={Dynamic malware analysis with feature engineering and feature learning},
  author={Zhang, Zhaoqi and Qi, Panpan and Wang, Wei},
  booktitle={Proceedings of the AAAI conference on artificial intelligence},
  volume={34},
  number={01},
  pages={1210--1217},
  year={2020}
}

@misc{audit,
  year = 2016,
  url = {https://linux.die.net/man/8/auditd/.},
  title = {The Linux audit daemon},
}

@article{xiong2020conan,
  title={CONAN: A practical real-time APT detection system with high accuracy and efficiency},
  author={Xiong, Chunlin and Zhu, Tiantian and Dong, Weihao and Ruan, Linqi and Yang, Runqing and Chen, Yan and Cheng, Yueqiang and Cheng, Shuai and Chen, Xutong},
  journal={IEEE Transactions on Dependable and Secure Computing},
  year={2020},
  publisher={IEEE}
}

@inproceedings{hossain2017sleuth,
  title={$\{$SLEUTH$\}$: Real-time attack scenario reconstruction from $\{$COTS$\}$ audit data},
  author={Hossain, Md Nahid and Milajerdi, Sadegh M and Wang, Junao and Eshete, Birhanu and Gjomemo, Rigel and Sekar, R and Stoller, Scott and Venkatakrishnan, VN},
  booktitle={USENIX Security Symposium },
  pages={487--504},
  year={2017}
}

@misc{darpa3,
  year = 2018,
  title = {Transparent Computing Engagement 3 Data Release},
  note={\url{https://github.com/darpa-i2o/TransparentComputing/blob/master/README-E3.md.}},
  author = {A. D. Keromytis},
}

@inproceedings{inam2022sok,
  title={SoK: History is a Vast Early Warning System: Auditing the Provenance of System Intrusions},
  author={Inam, Muhammad Adil and Chen, Yinfang and Goyal, Akul and Liu, Jason and Mink, Jaron and Michael, Noor and Gaur, Sneha and Bates, Adam and Hassan, Wajih Ul},
  booktitle={2023 IEEE Symposium on Security and Privacy (SP)},
  pages={307--325},
  year={2022},
  organization={IEEE Computer Society}
}

@inproceedings{kwon2018mci,
  title={MCI: Modeling-based Causality Inference in Audit Logging for Attack Investigation.},
  author={Kwon, Yonghwi and Wang, Fei and Wang, Weihang and Lee, Kyu Hyung and Lee, Wen-Chuan and Ma, Shiqing and Zhang, Xiangyu and Xu, Dongyan and Jha, Somesh and Ciocarlie, Gabriela F and others},
  booktitle={NDSS},
  volume={2},
  pages={4},
  year={2018}
}

@article{zhu2023aptshield,
  title={APTSHIELD: A Stable, Efficient and Real-time APT Detection System for Linux Hosts},
  author={Zhu, Tiantian and Yu, Jinkai and Xiong, Chunlin and Cheng, Wenrui and Yuan, Qixuan and Ying, Jie and Chen, Tieming and Zhang, Jiabo and Lv, Mingqi and Chen, Yan and others},
  journal={IEEE Transactions on Dependable and Secure Computing},
  year={2023},
  publisher={IEEE}
}

@inproceedings{tang2018nodemerge,
  title={Nodemerge: Template based efficient data reduction for big-data causality analysis},
  author={Tang, Yutao and Li, Ding and Li, Zhichun and Zhang, Mu and Jee, Kangkook and Xiao, Xusheng and Wu, Zhenyu and Rhee, Junghwan and Xu, Fengyuan and Li, Qun},
  booktitle={Proceedings of the 2018 ACM SIGSAC Conference on Computer and Communications Security},
  pages={1324--1337},
  year={2018}
}

@misc{etw,
  title = {About Event Tracing},
  year = 2018,
  url = {https://docs.microsoft.com/en-us/windows/win32/etw/about-event-tracing/.},
}

@inproceedings{ma2016protracer,
  title={Protracer: Towards Practical Provenance Tracing by Alternating Between Logging and Tainting.},
  author={Ma, Shiqing and Zhang, Xiangyu and Xu, Dongyan and others},
  booktitle={NDSS},
  volume={2},
  pages={4},
  year={2016}
}

@inproceedings{xu2022depcomm,
  title={Depcomm: Graph summarization on system audit logs for attack investigation},
  author={Xu, Zhiqiang and Fang, Pengcheng and Liu, Changlin and Xiao, Xusheng and Wen, Yu and Meng, Dan},
  booktitle={2022 IEEE Symposium on Security and Privacy (SP)},
  pages={540--557},
  year={2022},
  organization={IEEE}
}

\end{document}